# Probing GHz Spin Dynamics Across Magnetic Phase Transitions in $CrCl_3$ Nanoflakes Using Nitrogen-Vacancy Microscopy

Benjamin Hammons,[1&] Jitender Kumar,[1&*] Sehrish Iqbal,[1] Prem Bahadur Karki,[2] Karishma Prasad,[3] Tianlin Li,[4] Aram Pirali,[4] Ayodimeji E. Aregbesola,[2] Rupak Timalsina,[1] Xia Hong,[4] Jian Wang,[3] Kapildeb Ambal,[2] Ilja Fescenko,[5*] and Abdelghani Laraoui[1,4*]

[1]Department of Mechanical & Materials Engineering, University of Nebraska-Lincoln, Lincoln, NE 68588, United States
[2]Department of Mathematics, Statistics, and Physics, Wichita State University, Wichita, KS 67260, United States
[3]Department of Chemistry and Biochemistry, Wichita State University, Wichita, KS 67260, United States
[4]Department of Physics and Astronomy and Nebraska Center for Materials and Nanoscience, University of Nebraska-Lincoln, Lincoln, NE 68588, United States
[5]Laser Center, Faculty of Science and Technology, University of Latvia, LV-1004 Riga, Latvia
[&] Equal Contribution
[*]Corresponding Authors: jkumar5@unl.edu, ilja.fescenko@lu.lv, alaraoui2@unl.edu

## Abstract

$CrCl_3$, a layered van der Waals (vdW) magnet, exhibits in-plane magnetic anisotropy and enhanced interlayer coupling upon stacking, making it an ideal platform to host exotic nanoscale magnetic phenomena such as magnon hydrodynamics and meron-like topological spin defects. When interfaced with other vdW materials, its antiferromagnetic-to-ferromagnetic and ferromagnetic-to-paramagnetic phase transitions and magnetic anisotropy can be tuned by voltage, strain, and layer stacking. Understanding the spin dynamics of $CrCl_3$ at its magnetic phase transitions is crucial to its applications in magnonics. Here, we investigate the spin dynamics of $CrCl_3$ nanoflakes using cryogenic diamond quantum sensing microscopy, based on measuring optically detected magnetic resonance, Rabi oscillations, and spin-lattice relaxation time ($T_1$) of shallow nitrogen vacancy (NV) centers in diamond. In the ferromagnetic regime, we observe a pronounced reduction in the NV spin resonance contrast, a collapse of the Rabi oscillations, and a strong enhancement by two orders of magnitude of the relaxation rate $\Gamma_1 = 1/T_1$. These observations indicate intensified spin fluctuations in the gigahertz range. Broadband ferromagnetic resonance spectroscopy on $CrCl_3$ crystals reveals resonance frequencies in the 4–15 GHz range together with a linewidth of ~24 mT, further supporting the NV measurements. A phenomenological model of magnetic-noise-induced NV relaxation reproduces the temperature dependence of $\Gamma_1$ by combining antiferromagnetic, ferromagnetic, and paramagnetic fluctuation channels, indicating that magnetic noise is strongest in the ferromagnetic regime and evolves markedly across the phase transition. These results are crucial for using $CrCl_3$ in 2D magnonics and hybrid quantum-magnon systems.

## Introduction

Two-dimensional (2D) van der Waals (vdW) magnets have come to the forefront of condensed matter physics research in recent years for their highly tunable magnetic order and potential for developing nanoscale spintronics and magnonics.[1–4] Strong magnetic anisotropy and weak interlayer coupling enable these materials to exhibit a variety of intriguing properties, ranging from layer-dependent magnetic order (antiferromagnetic or ferromagnetic) and multiferroicity to giant magnetoresistance and superconductivity.[5–8] A notable family of 2D vdW magnetic system is the family of chromium trihalides, Cr*X₃* (X = Cl, Br, I), where ferromagnetism in a truly 2D monolayer was first discovered in $CrI_3$.[9] Layered $CrI_3$ and $CrBr_3$ order ferromagnetically below 68 K and 33 K, respectively.[9,10] In $CrCl_3$, the $Cr^{3+}$ ions with spin $S = 3/2$ are arranged ferromagnetically within each layer and antiferromagnetically between adjacent layers, see Figure 1a.[11] Upon cooling, $CrCl_3$ exhibits two magnetic phase transitions within a narrow temperature window: a transition from the paramagnetic (PM) to the ferromagnetic (FM) phase at ~18 K, followed by a transition from the FM to the antiferromagnetic (AFM) phase below ~13 K.[11,12] In addition, $CrCl_3$ exhibits rich spin dynamics. For example, the AFM resonance frequencies of the optical and acoustic modes lie in the GHz regime due to weak interlayer coupling,[13] making it promising for antiferromagnetic spintronics.[14,15] Interestingly, beyond conventional magnons, hydrodynamic-like magnon flow and a low-energy two-dimensional magnon sound mode in the MHz regime have been observed in the ferromagnetic state of monolayer and layered $CrCl_3$.[16] Unlike $CrBr_3$ and $CrI_3$, gapless Dirac magnons have been reported in $CrCl_3$ in the THz frequency range, originating from FM interactions on the honeycomb lattice.[17,18] These results indicate that spin dynamics in vdW $CrCl_3$ span a broad frequency range (MHz – THz ), making $CrCl_3$ an excellent candidate for 2D magnonics.[19–22]

Despite the direct observation of coherent magnons in bulk $CrCl_3$ crystals[23] and magnetic transport studies of ultrathin $CrCl_3$ flake-based magnetic tunnel junctions,[24,25] measurements of the intrinsic spin fluctuations of thin vdW $CrCl_3$ nanoflakes, particularly in the GHz regime, are still lacking.[13–15,17,18] In modern spintronics, the magnetic noise profile of a system plays a vital role in device development and strongly influences device density, scalability, and stability.[26–29] It has been shown that magnon dynamics, magnon scattering, and magnon–phonon interactions associated with spin and magnon fluctuations leave distinct signatures in the magnetic noise spectrum.[30–32] Conventional techniques such as superconducting quantum interference devices (SQUID),[33,34] inelastic neutron scattering,[17] and ferromagnetic resonance (FMR)[34,35] are primarily designed to characterize static and/or coherent dynamic magnetic properties on bulk crystals and are generally unable to access non-coherent magnetic fluctuations over a broad frequency range at the nanoscale. Recently, nitrogen vacancy (NV) centers in diamonds have emerged as highly sensitive quantum sensors capable of detecting not only DC and AC magnetic fields but also spin fluctuations over a large frequency range.[22,36–42] For instance, NV optically detected magnetic resonance (ODMR), Rabi oscillations, and transverse (spin-spin) coherence times ($T_2^*$ and $T_2$) are sensitive to magnetic noise from the DC to MHz regime, while the longitudinal spin-lattice relaxation time ($T_1$) is affected by spin fluctuations in the GHz range.[36,38] Equipped with these capabilities and an excellent spatial resolution (≤ 40 nm), NV microscopy/spectroscopy has become a powerful approach for probing both coherently and non-coherently coupled magnons in various magnetic materials.[22,36,38,43–45]

Recently, NV $T_2$ spectroscopy was used to study the MHz noise generated by magnon sound modes in $CrCl_3$ monolayers/multilayers.[16] In another study, ODMR spectroscopy of NV-contained nanodiamonds was used to probe coherent antiferromagnetic resonances of bulk $CrCl_3$ crystals.[13]

However, the study of GHz spin dynamics across multiple magnetic phases in $CrCl_3$ flakes is not realized yet.

In this work, we performed cryogenic (4.3 – 45 K) ODMR, Rabi oscillations, and $T_1$ relaxometry measurements of shallow (~ 6 nm) NV centers across the magnetic phase transitions of $CrCl_3$ nanoflakes transferred onto the diamond surface. The distinct magnetic phases of $CrCl_3$ are clearly resolved, evident from the pronounced reduction of the ODMR contrast and collapse of the Rabi oscillation (decrease in amplitude and decay time) in the FM phase. A significant increase in the NV relaxation rate $\Gamma_1 = 1/T_1$ was measured on the $CrCl_3$ nanoflakes; notably, the temperature dependence of $\Gamma_1$ is higher in the FM phase than in the PM and AFM states. Magnetic field dependence measurements confirm the modification of the spin-wave bath particularly in the FM region, explained by the existence of short-range ferromagnetic correlations. We further performed low-temperature FMR spectroscopy on $CrCl_3$ crystals, and the results are consistent with the NV measurements, confirming the increased FM magnon excitations in the 4–15 GHz regime with a linewidth of ~24 mT.

## Results and Discussion

**NV magnetometry of $CrCl_3$ nanoflakes.** Bulk $CrCl_3$ crystals were mechanically exfoliated and transferred to diamond substrates implanted with a shallow (~6 nm) NV layer.[22,40,46] See Supporting Information Section S1 for more details. Large flakes with a thickness of 42–60 nm were identified by optical microscopy and atomic force microscopy, and their spin dynamics were measured using cryogenic widefield diamond quantum sensing microscopy.

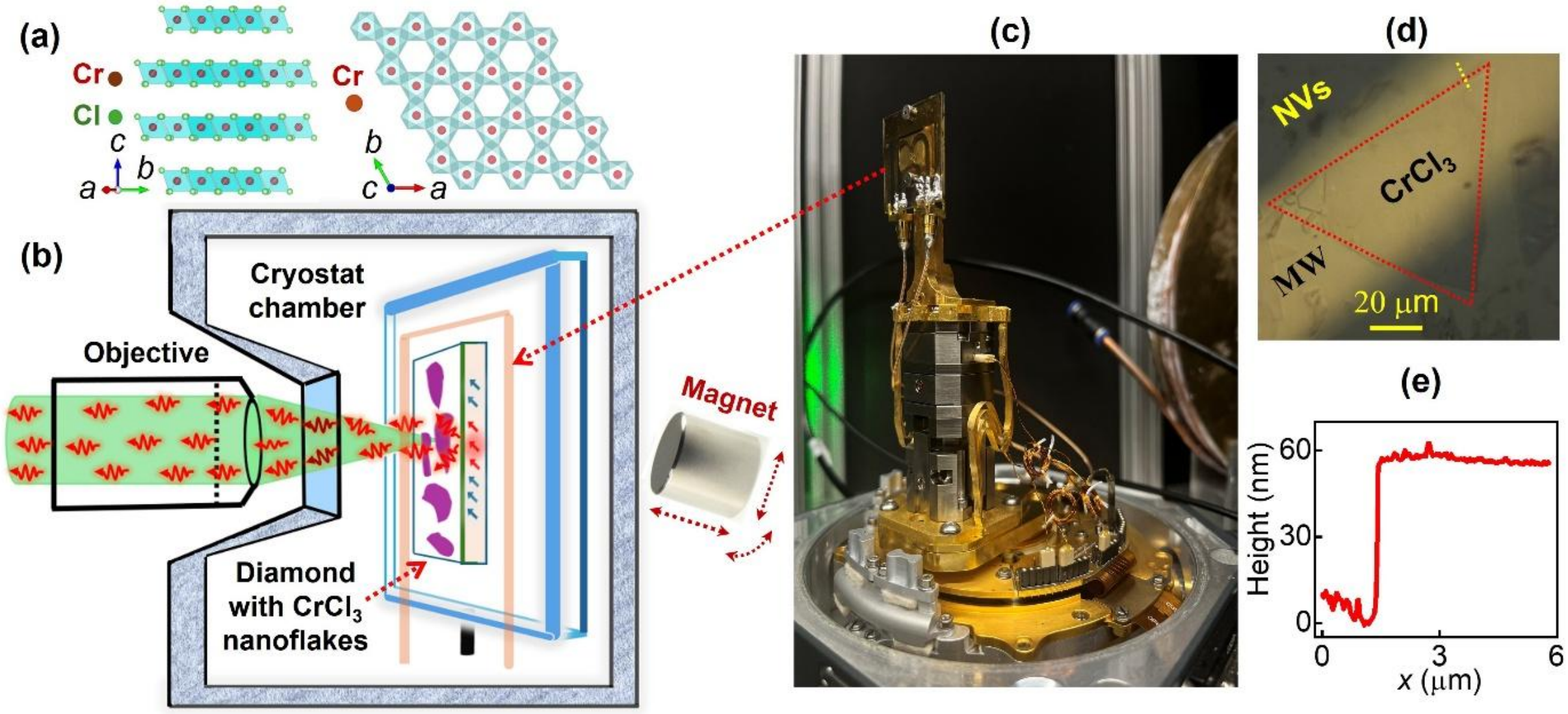

**Figure 1.** (a) Crystal structure of $CrCl_3$ in the R$\overline{3}$ phase (side view) and the honeycomb lattice of Cr spins in a single layer of $CrCl_3$ (top view). Cl atoms are omitted for clarity. Schematic (b) and a photograph (c) of the cryogenic diamond quantum sensing microscope. It is housed inside the cryostat chamber; an objective is outside of the chamber to focus the green (532 nm) laser and collect the NV red (650-710 nm) fluorescence; $CrCl_3$ flakes are transferred onto the diamond, which is mounted on a sapphire substrate with MW stripline; the sample stage (sapphire base with microwave (MW) line, and the mounted diamond) is controlled by Attocube (*X, Y, Z*) piezo-motors for precise motion. A permanent magnet is used outside the chamber to apply magnetic field. (d) Optical image of the $CrCl_3$ flake transferred on the diamond. (e) Atomic force microscopy height profile of the flake in (d), showing a flake height of ~60 ± 1 nm.

In Figures 1b and 1c, we show the schematic and the picture of our cryogenic widefield NV microscope used to study the spin dynamics of $CrCl_3$ nanoflakes, respectively. The cryogenic chamber, sapphire base with microwave (MW) line, and the mounted diamond are depicted in Figure 1b. For improved thermal contact with the cold finger of the closed-cycle refrigerator (CCR), the diamond was attached to a sapphire substrate using VGE varnish. The sapphire substrate (25 mm × 25 mm) is patterned with MW antennas that are soldered to the MW SMA connectors.

A green (532 nm) laser is used to excite the NV spins over an area of ~ 23 × 23 μm$^2$ and the red (650–710 nm) fluorescence is detected using an avalanche photo detector (APD) and a sCMOS camera.[39,40,47] We used an electronic grade diamond (2 × 1 × 0.1 mm$^3$) cut and polished along the (100) plane. The 6 nm deep NV layer is formed underneath the diamond surface by ion-beam implantation of $^{15}$N ions, followed by high temperature (1373 K for 2 hours) and high-vacuum (2 × 10$^{-6}$ Torr) annealing, and cleaning in a boiling (473 K) tri-acid mixture to remove graphite residues.[48–50] The diamond substrate with $CrCl_3$ nanoflakes is placed on the sapphire coverslip, patterned with gold (Au) striplines for NV spin manipulation, see Figure 1d. Due to the high autofluorescence coming from the $CrCl_3$ flakes for wavelengths > 720 nm (see Section S2 in Supporting Information), we used a bandpass filter to detect the NV fluorescence below 710 nm.

Figure 2a depicts NV ODMR spectra for a temperature window (4.7 – 28 K) that covers the magnetic phases of the $CrCl_3$ flake (thickness of ~60 ± 1 nm). We used low MW (~50 mW) and laser (~36 mW) powers to prevent excess heating and flake degradation. Four peaks appear in the ODMR spectrum: $f_-$, $f_+$ for NV spins aligned along [111] and $f_{m-}$, $f_{m+}$ for NV spins merged and aligned in the opposite direction of the applied magnetic field $H$ of 1.45 mT.[22] A clear ODMR contrast decrease is observed for all NV peaks in the AFM and FM states. Figure 2b shows the zoomed temperature-dependent ODMR spectra of $f_-$ in Figure 2a. Interestingly, the normalized fluorescence contrast clearly separated into three different bands: AFM, FM, and PM. The separation temperatures are well matched with the magnetic phase transitions reported for $CrCl_3$.[11] In CW-ODMR, the normalized NV fluorescence contrast depends on the population difference between $m_S = |0\rangle$ and $m_S = |\pm 1\rangle$ electronic spin states.[51] This population difference is modified when the external MW field $B_1$ couples to the NV spins and drives transitions between these states.[51] The $f_-$ ODMR spectrum recorded at different temperatures (4.7 – 28 K) is fitted with a Lorentzian to extract the change in contrast, area, and linewidth. The fitting results corresponding to the $f_-$ peak are shown in Section S3, Figures S3.1 and S3.2 in the Supporting Information. The signal-to-noise ratio (SNR) in our measurements is ~41.66 for an averaging time of 1 second.

The resulting ODMR contrast and area are plotted as a function of temperature in Figures 2c and 2d, respectively. A sharp reduction of both contrast and area is observed when $CrCl_3$ enters the FM regime (13-18 K). Upon further lowering the temperature when the system transits from FM state to AFM state at a temperature of ~12 K, the ODMR contrast/area recovers, as in the case of the PM state. This behavior suggests enhanced magnon-mediated MW absorption and increased magnetic noise in the FM state. To further showcase these effects, we plotted the 2D in-situ (frequency vs temperature) map of all ODMR peaks (Figure 2e) and zoomed $f_-$ peak (Figure 2f), showing clear AFM-to-FM and FM-to-PM transitions. Very recently, an almost complete collapse of CW-ODMR contrast was observed in the 2D ferromagnet $Fe_3GeTe_2$ in its FM state, where the contrast reduction is attributed to quasi-static noise arising from magnetic domains reversal.[52]

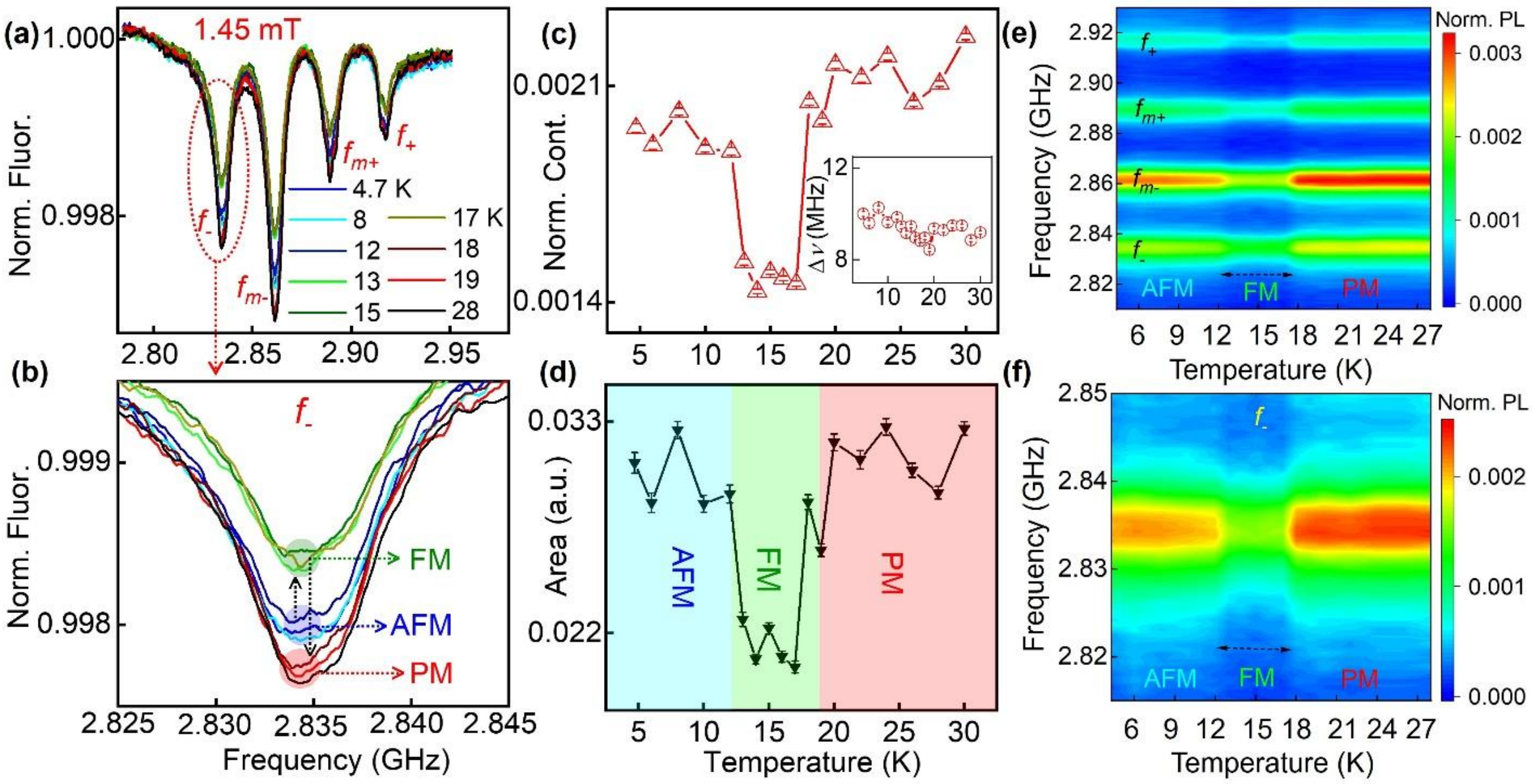

**Figure 2.** (a) CW-ODMR spectra of ensemble NVs on the $CrCl_3$ flake placed on top of the diamond, measured at different temperatures (4.7 – 28 K). Four distinct ODMR dips ($f_-$, $f_{m-}$, $f_+$, and $f_{m+}$) are observed at an external magnetic field of 1.45 mT applied along the NV quantization axis. (b) A zoomed-in view of the $f_-$ ODMR peak, showing clear distinctions among the three magnetic phases (AFM, FM, PM), visible as abrupt changes in contrast across the phase transitions. ODMR contrast (c) and area (d), extracted from the Lorentzian fitting of the $f_-$ ODMR peak, versus temperature. The inset in (c) shows the corresponding change in full width at half maximum (FWHM) vs temperature. 2D in-situ (frequency vs temperature) map of all ODMR peaks (e) and zoomed $f_-$ peak (f), showing clear AFM-FM and FM-PM transitions.

To further elucidate the ODMR changes in the FM state of the $CrCl_3$ flake, we measured NV Rabi oscillations (NV fluorescence vs MW pulse duration) on and off the $CrCl_3$ flake versus temperature in the range of 4.3 – 45 K and at various applied magnetic field amplitudes (1.45 – 40.38 mT). The Rabi measurements were performed on $f_-$ at high (~10 W) MW and low (72 mW) laser powers. Figure 3a displays NV Rabi oscillations at $H$ of 12.56 mT for selected temperatures (6.3, 14, 18, 21, and 26 K) spanning the AFM, FM and PM states (top to bottom). A temperature increase of ~2 K is deduced from the shift of the dip position in the temperature-dependent signal relative to ODMR and $T_1$ measurements discussed below. For ODMR measurements, the MW duty cycle is 100% leading to a MW power of ~ 50 mW. While for Rabi measurements, the MW duty cycle is ~2.5%, giving a MW power of ~250 mW, which is 5 times higher than ODMR.

A near-complete suppression of the Rabi oscillations in the FM state was obtained. The Rabi signal revives in the AFM and PM phases. This behavior is clearly evident in the frequency domain (see Figure 3b), where the Fourier transform of the time-domain Rabi signal reveals the corresponding disappearance of the amplitudes in the FM phase, further confirming the loss of coherent MW spin driving in the FM phase window.

For the quantitative estimation of the Rabi amplitude and decay time, we have fitted the Rabi oscillations with equation Eq.1 as:

$$f(\tau) = e^{-\tau/T_{Rabi}}(A_c \cos(\omega_{Rabi}\tau) + A_s \sin(\omega_{Rabi}\tau)) + B, \qquad \text{Eq.1}$$

where $T_{Rabi}$ is the Rabi decay time, $\omega_{Rabi}$ is the Rabi frequency, $B$ is the constant offset. The two quadrature amplitudes $A_c$ and $A_s$ are fitted independently. From them, the Rabi oscillation amplitude related to the NV contrast is obtained as $A = \sqrt{A_c^2 + A_s^2}$, and the oscillation phase $p = \mathrm{atan2}(A_s, A_c)$. The phase $p$ is therefore not a free fit parameter but is derived from the fitted amplitudes. Figures 3c and 3d show the temperature dependency of the Rabi amplitude and $T_{Rabi}$ obtained from the fitting parameters for on and off flake at $H$ of 1.45 and 12.56 mT. The pronounced reduction of both Rabi amplitude and decay in the FM phase can be explained by many effects including the enhanced magnetic fluctuations arising from FM domains and/or by the increased population of spin waves (magnons) that generates a strong noise in the rotating frame that accelerates the rapid decay of $T_{Rabi}$ time. To confirm the intrinsic origin of these effects, control measurements were performed without the $CrCl_3$ flake. The Rabi amplitude/decay obtained at $H$ of 1.45 mT off the flake (insets of Figures 3c and 3d) shows no features across the AFM, FM, and PM phases, confirming that the observed decoherence in the FM region originates from the $CrCl_3$ flake. Additional Rabi measurements at different magnetic fields (1.45, 12.56, 24.51, and 40.38 mT), together with their corresponding fits are provided in the Supporting Information Section S3. At higher magnetic fields, the FM phase window shifts toward higher temperatures, consistent with the field-supported stabilization of the magnetic order in $CrCl_3$.[12]

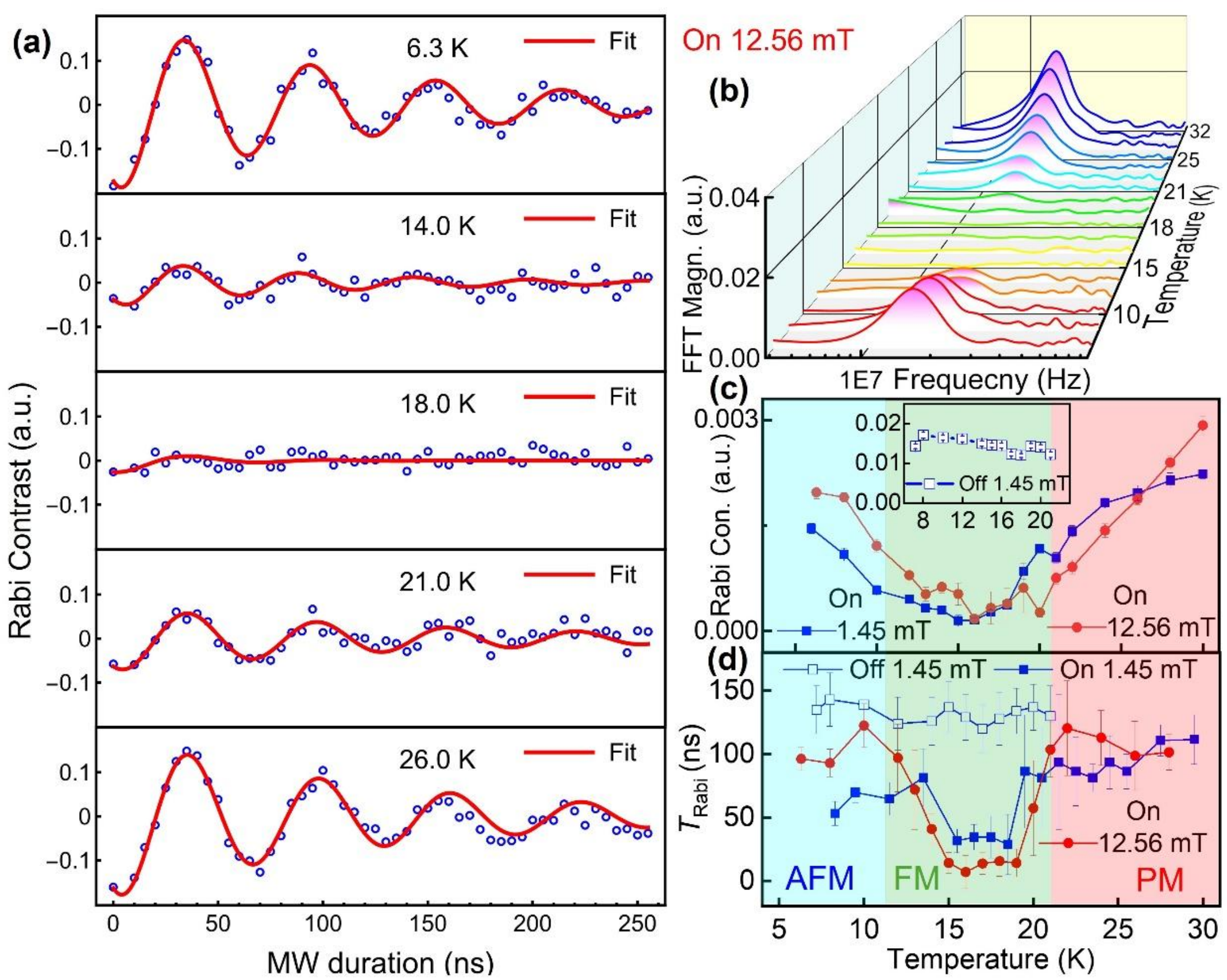

**Figure 3.** (a) Rabi oscillations (fluorescence vs time) of NVs underneath the $CrCl_3$ flake recorded at the three different magnetic state temperature regimes: PM (21 K, 26 K), FM (14 K, 18 K), and AFM (6.3 K). (b) Fast Fourier Transformation (FFT) of the Rabi oscillations in (a) and Figure S3.3.1 in the Supporting Information, recorded at temperatures in the range 6.3 – 32 K. Rabi amplitude (c) and decay $T_{\mathrm{Rabi}}$ (d) versus temperature measured at $H$ of 1.45 mT (filled rectangles) and 12.56 mT (filled circles). The control (off the flake) data at $H$ of 1.45 mT are plotted (open rectangles) in the insets of (c) and (d) for the Rabi amplitude and decay, respectively.

The temperature dependence of the Rabi decay time $T_{Rabi}$ at higher magnetic fields (e.g., 40.38 mT, Figure S3.6 in the Supporting Information) shows an extended region of suppressed NV coherence toward elevated temperatures. Notably, at these higher fields, the Rabi oscillations become increasingly difficult to fit reliably even up to high temperatures (> 25 K). We ascribe this characteristic to the extension of FM correlations beyond the low field phase boundary, leading to enhanced magnetic noise.[12]

We also performed Rabi measurements on a 42 nm thick $CrCl_3$ flake, which was not on the top of the MW line as in the case of the 60 nm thick flake. The measured Rabi signals at a magnetic field of 27.77 mT versus temperature are shown in Supporting Information Section S4 off (Figure S4.3) and on (Figure S4.5) the $CrCl_3$ flake (Supporting Information). The Rabi amplitude and decay are minimum in the ordered magnetic state (AFM, FM), see Figure S4.1 in the Supporting Information. Interestingly, Rabi oscillations do not collapse, suggesting a weaker magnon population induced by the low MW intensity far away (~20 μm) from the MW stripline line.

We note there is no change in the $B_1$ field in the FM regime (see Figure S3.3.2 in the Supporting Information) suggesting that the observed effects on Rabi oscillations are related to the increased decoherence induced by the ferromagnetic spin noise, confirmed by $\Gamma_1$ measurements.

**NV spin-noise spectroscopy of spin dynamics in $CrCl_3$ nanoflakes.** Despite the AFM ground state of $CrCl_3$ at low temperatures, magnon frequencies (optical and acoustic) extend down to the GHz regime due to the material's weak magnetic anisotropy.[13] Conventional magnetic resonance techniques such as FMR and antiferromagnetic resonance (AFMR) techniques have been used to detect spin dynamics in bulk $CrCl_3$ crystals.[13,23] However, they lack sensitivity to measure thin-layered $CrCl_3$ flakes. NV spectroscopy is a remarkably powerful technique for studying spin dynamics at the nanoscale, as it directly maps the magnetic noise and/or the stray field produced by spin waves of the magnet.[22,36,38] In particular, the NV spin lattice relaxation time $T_1$ is an excellent tool for detecting these GHz-scale fluctuations, due to its zero-field splitting (ZFS) in the GHz regime (~2.87 GHz).[40,46,53] The NV relaxation rate $\Gamma_1 = 1/T_1$ is susceptible to surrounding magnon noise bath perpendicular to the NV quantization axis and accelerates in the presence of external noise as:[40,46,54]

$$\Gamma_1 = \Gamma_{int} + \gamma_{NV}^2 \langle B^2 \rangle \int S(\omega, T, E_a) F_1(\omega) d\omega, \qquad \text{Eq.2}$$

where $\gamma_{NV}$ is the NV electron spin gyromagnetic ratio, $\sqrt{\langle B^2 \rangle}$ is the effective magnetic field at the NV position, $\Gamma_{int} = 1/T_{1,int}$ is the intrinsic NV relaxation rate, and $S(\omega, T, E_a)$ is the noise spectral density of the noise bath as function of frequency (ω) and temperature (T) and magnetic anisotropy energy ($E_a$), and $F_1(\omega)$ is the filter for $T_1$ measurement: [40,55]

$$F_1(\omega) = \sum_{i=\pm 1} \left[ \frac{4\pi / T_2^*}{(2\pi / T_2^*)^2 + (\omega - \omega_i)^2} \right], \qquad \text{Eq.3}$$

where $T_2^*$ is the NV dephasing (Ramsey) time and $\omega_{i=\pm 1}$ is the NV spin resonance frequencies for $m_s = 0 \rightarrow m_s = \pm 1$. The $T_1$ measurements were performed on $f_-$ at a high (~10 W) MW power with an MW duty cycle of ~0.4% for $\tau$ = 20 μs, giving a MW power of ~40 mW. This value is comparable to ODMR and is six times less than Rabi measurements.

Figure 4a shows $T_1$ (fluorescence vs time) curves taken on and off the 60-nm thick $CrCl_3$ flake at $H$ of 1.45 mT and three temperatures (4.3, 15, and 30 K), representing the distinct magnetic (AFM, FM, PM) phases. The NV $T_1$ curves are fitted with an exponential decay to deduce $T_1$:

$$S = \tau_0 e^{\left(-\frac{\tau}{T_1}\right)^n}, \qquad \text{Eq.4}$$

where $S$ is the fluorescence intensity, $\tau_0$ is the prefactor, $\tau$ is the waiting time, and $n$ is the stretching parameter. Interestingly, $T_1$ time in the presence of the flake is drastically reduced by two orders of magnitude. For instance, the "off-flake" $T_1^{OFF}$ is 1.19 ± 0.05 ms at 30 K, whereas the "on-flake" $T_1^{PM}$ at the same temperature (PM phase) is 25.9 ± 1.6 μs. This represents a $\frac{T_1^{OFF}}{T_1^{PM}}$ ~ 45-fold reduction due to the PM Cr spin noise, a result that closely matches the 35-fold reduction recorded in the PM phase of α-$RuCl_3$.[56] $T_1$ is further reduced as the temperature lowered and reaching $\frac{T_1^{OFF}}{T_1^{FM}} \sim 120$ in the FM phase at 15 K, similar to those effects from $Sr_2FeReO_6$ flakes on NV $T_1$ around $T_C$.[57] However, in AFM phase at 4.3 K, the system shows a mild recovery $\frac{T_1^{OFF}}{T_1^{AFM}} \sim 64$. Similar trends were observed from temperature dependence of electron spin resonance (ESR) and FMR on $CrCl_3$ crystals.[58]

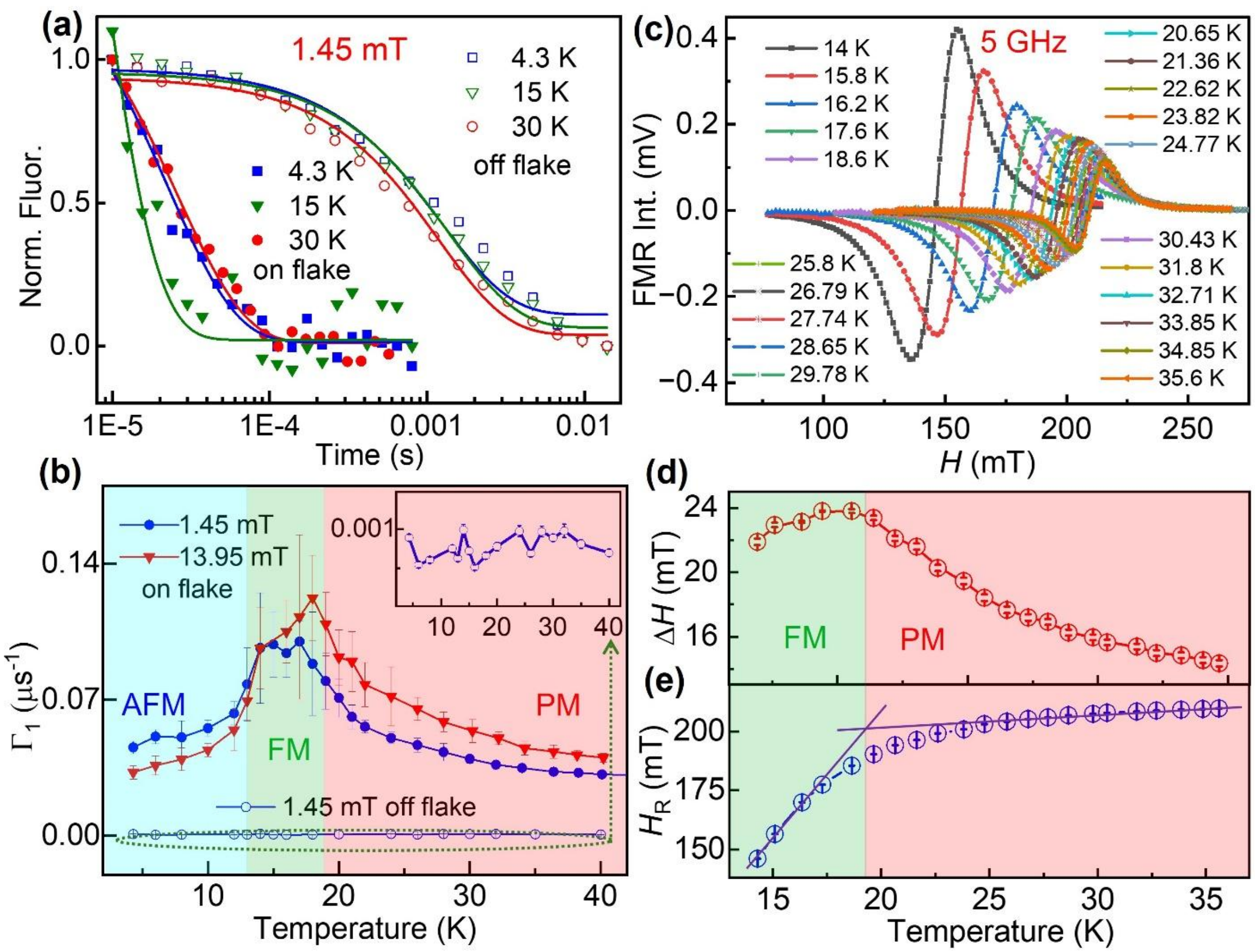

**Figure 4.** (a) Measured (scattered) and fitted (solid lines) $T_1$ (fluorescence vs time) curves at three temperatures (4.3, 15, and 30 K) on and off the 60 nm $CrCl_3$ flake. (b) $\Gamma_1$ vs temperature at $H$ of 1.45 and 13.95 mT on the flake. The off flake $\Gamma_1$ curve is zoomed in the inset for better presentation, where no effect is observed. (c) Temperature-dependent FMR (intensity vs applied magnetic field $H$) spectra of $CrCl_3$ crystals measured at a MW frequency of 5 GHz, recorded at temperatures in the range of 13.9 – 35.6 K. Temperature dependence of the linewidth $\Delta H$ (d) and resonance field $H_R$ (e), obtained by fitting the FMR spectra in (c) with a Lorentzian function (see Supporting Information Section S5).

In $CrCl_3$, the magnetic phase transitions are closely spaced, and the transition from the PM to the FM phase is second-order in nature, making the material critical near $T_C$, and spin correlations diverge in both space and time. This acceleration of magnetic fluctuations, known as critical slowing down, imprints onto the imaginary part of the dynamical magnetic susceptibility, $\mathrm{Im}\chi(\omega)$, according to the fluctuation–dissipation theorem.[36,59,60] The NV $\Gamma_1$ relates to the imaginary magnetic susceptibility as:[36,60,61]

$$\Gamma_1 = \frac{\mu_B^2}{2\hbar}\coth\left(\frac{\beta\hbar\omega}{2}\right)\mathrm{Im}\chi_{B_zB_z}(z,\omega)\ , \qquad \text{Eq.5}$$

where $\mu_B$ is Bohr magneton, $\hbar = h/2\pi$ ($h$ is Planck constant), $z$ is the NV standoff, $\omega$ is the NV spin resonance frequency, $\beta = \frac{1}{k_B T}$ ($k_B$ is the Boltzmann constant and $T$ is temperature), and $B_z$ is the magnetic fluctuations field at point $z$.

To further explore this behavior, we present the temperature-dependent $\Gamma_1$ with close temperature steps in Figure 4b at two different magnetic fields of 1.45 mT and 13.95 mT. The measured/fitted NV $T_1$ curves for each temperature are displayed in the Supporting Information Section S3. The $\Gamma_1$ curves show a clear peak at ~18-19 K related to the PM-to-FM phase transition, and a shoulder-like feature at ~12-13 K related to the FM-to-AFM phase transition. The strong peak in $\Gamma_1$ at FM-PM transition is in line with the published imaginary part of magnetization ($m''$) at FM-PM transition of bulk $CrCl_3$ crystals.[12]

The intrinsic spin dynamics of NV centers in the absence of the flake remain featureless across the magnetic phase transition temperatures regime. This confirms that the observed sharp features in $\Gamma_1$ are directly coupled to the evolution of spin dynamics across the magnetic phase transitions in $CrCl_3$. In $CrCl_3$, the second-order phase transition (PM to FM) is susceptible to non-thermal parameters, including pressure and external magnetic fields.[12,62] To investigate this, we explored the effect of a higher external magnetic field (13.95 – 40.38 mT) on the NV $\Gamma_1$ profile of the $CrCl_3$ flake, as shown in Figure 4b, and Figures S3.6c and S3.6f in the Supporting Information. A clear enhancement in $\Gamma_1$, extended from ~13 K to elevated temperatures (up to 24 K) is observed, indicating that the magnetic field stabilizes magnetic correlations further into the paramagnetic regime, a typical feature of the 2D magnetism in $CrCl_3$.[63] This finding aligns with the existence of short-range ferromagnetic correlations previously identified via high-field ESR and FMR spectroscopy in bulk $CrCl_3$ crystals.[58] Furthermore, these effects were reproduced using a second $CrCl_3$ flake of thickness ~42 nm, located far away (~20 μm) from the MW line, demonstrating excellent consistency across flakes. These measurements were discussed in Supporting Information Section S4 and Figure S4.2.

To elucidate the origin of spin noise detected in the NV measurements (Rabi and $\Gamma_1$), we performed temperature-dependent FMR measurements on $CrCl_3$ crystals (size ~ a few mm) in a MW frequency range (4 – 15 GHz), see Methods. Figure 4c shows the measured magnetic resonance spectra conducted at a 5 GHz resonance frequency from ~13.9 to 35.6 K. We fit each magnetic resonance spectrum using a derivative of a single Lorentzian function.[23,64] From the fit parameter, we extract the resonance linewidth ($\Delta H$) and resonance field ($H_R$). We repeated the measurements at various resonance frequencies at 7 GHz and 10 GHz, see Figure S5.2 in the Supporting Information. All three resonance frequencies exhibit similar temperature-dependent trends in both the linewidth and resonance fields. At higher temperatures (in the PM phase), the Cr spins are weakly coupled, and the FMR linewidth is determined by paramagnetic interactions leading to an ESR-like signal.[58] As the temperature decreases, the linewidth increases due to the onset of FM ordering, in which individual spins begin to undergo collective magnetic excitations (precessions) driven by the exchange interactions. $\Delta H$ reaches a maximum of ~ 24 mT at 16 –18 K (Figure 4d), corresponding to a damping constant $\alpha$ of ~$1 \times 10^{-3}$ (see the inset of Figure S5.1b, Supporting Information). As the system approaches the PM to FM phase transition $T_C$, strong critical spin fluctuations and magnetic inhomogeneity result in a maximum in the FMR linewidth. Near $T_C$, critical slowing down leads to strong dephasing of the uniform precession mode. Upon entering the ferromagnetic phase, the suppression of spin fluctuations and increased magnetic uniformity reduce damping, resulting in a narrower linewidth. Due to the low temperature limitation (13.9 K) of our FMR setup, we focused only on the FM and PM regions. $H_R$ increases linearly with temperature in the FM phase and saturates at ~209 mT in the PM phase. The intersection between the two lines in Figure 4e, gives a Curie temperature $T_C$ of ~19 K, in a good agreement with NV $\Gamma_1$ measurement. Indeed, the measured FMR resonance frequency vs $H_R$ curves at temperatures in the FM phase is fitted well with the Kittel formula for FMR (Figure S5.1b, Supporting Information). However, in the PM phase (Figure S5.1d, Supporting Information), they are fitted perfectly with a linear fit, indicative of paramagnetism (ESR-like signal).

To reflect the crossover between magnetic phases of $CrCl_3$ on the NV $\Gamma_1$, we developed a toy model of magnetic-noise-induced NV relaxation. In this model, the temperature dependence of $\Gamma_1$ is represented as a sum of phenomenological magnetic-noise channels associated with the AFM, FM, and PM phases, together with a slowly varying background contribution. Each channel is weighted by a smooth temperature window to account for the crossover between regimes. The model is formulated in $\Gamma_1$-space, where an additive decomposition is more natural. The total relaxation rate is written as:

$$\Gamma_1 (T) = \Gamma_{bg} (T) + \Gamma_0 (T) + \Gamma_{AFM} (T) + \Gamma_{FM} (T) + \Gamma_{PM} (T), \qquad \text{Eq.6}$$

where $\Gamma_{bg} (T)$ is a fixed linear background taken from independent measurements (in the temperature window of magnetic transitions), and $\Gamma_0 (T)$ is an additional weak, manually chosen linear contribution representing unresolved field-dependent relaxation not captured by the explicit magnetic channels.

The AFM contribution ($\Gamma_{AFM} (T)$) is modeled using a gapped magnon dispersion,[65] $\omega_{AFM}(k) = \sqrt{\Delta_A^2 + (v_A k^2)}$, with fixed gap $\Delta_A$ and fitted amplitude and velocity parameters. The FM contribution is modeled with quadratic Kittel magnon dispersion,[34] $\omega_{FM}(k) = \Delta_{F+}D_F k^2$, where $\Delta_F$ and $D_F$ are fitted with effective parameters. Both AFM and FM channels are evaluated through the same spectral integral for NV relaxation, which includes the NV-flake geometry, the NV depth, flake thickness, angular factor, and a temperature-dependent linewidth $\Delta H(T)$ obtained from FMR measurements discussed above. The linewidth enters the Lorentzian spectral denominator and sets the effective broadening of magnetic fluctuations. The PM contribution is described phenomenologically as $\Gamma_{PM}(T) \propto T\chi_{CW}(T)S(T)$, where $\chi_{CW}$ is a Curie-Weiss-like susceptibility term[66] and $S(T)$ is a Debye-type spectral factor built from a temperature-dependent correlation time: $\tau_c(T) = \tau_0 / [1 + (T|T_0)^n]$. Here $\tau_0$, $T_0$, and $n$ are effective fit parameters.

The AFM contribution is fitted at low temperature and then smoothly continued beneath the FM region up to the upper crossover temperature, see Figures 5a and 5b. The FM contribution is restricted to the intermediate temperature window. The PM contribution is activated above the upper crossover. These windows are not intended to represent microscopic phase fractions; they are only a smooth interpolation scheme that allows the three effective channels to be combined without discontinuities.

This decomposition highlights the coexistence of distinct magnetic fluctuation channels in the NV relaxation. At low temperature, $\Gamma_1$ is governed predominantly by AFM-associated fluctuations, which provide the main contribution to the gradual increase of $\Gamma_1$ on cooling toward the ordered state. In the intermediate-temperature range, however, this AFM-like contribution alone is not sufficient to reproduce the pronounced excess relaxation observed around the crossover region. To account for this additional spectral weight, the model requires a separate FM-like channel, which captures the enhanced magnetic noise appearing in the vicinity of the transition window. At higher temperatures, the localized FM-like contribution rapidly vanishes, while the remaining relaxation is described by a PM component with a broad high-temperature tail, consistent with thermally driven spin fluctuations above the ordered regime.

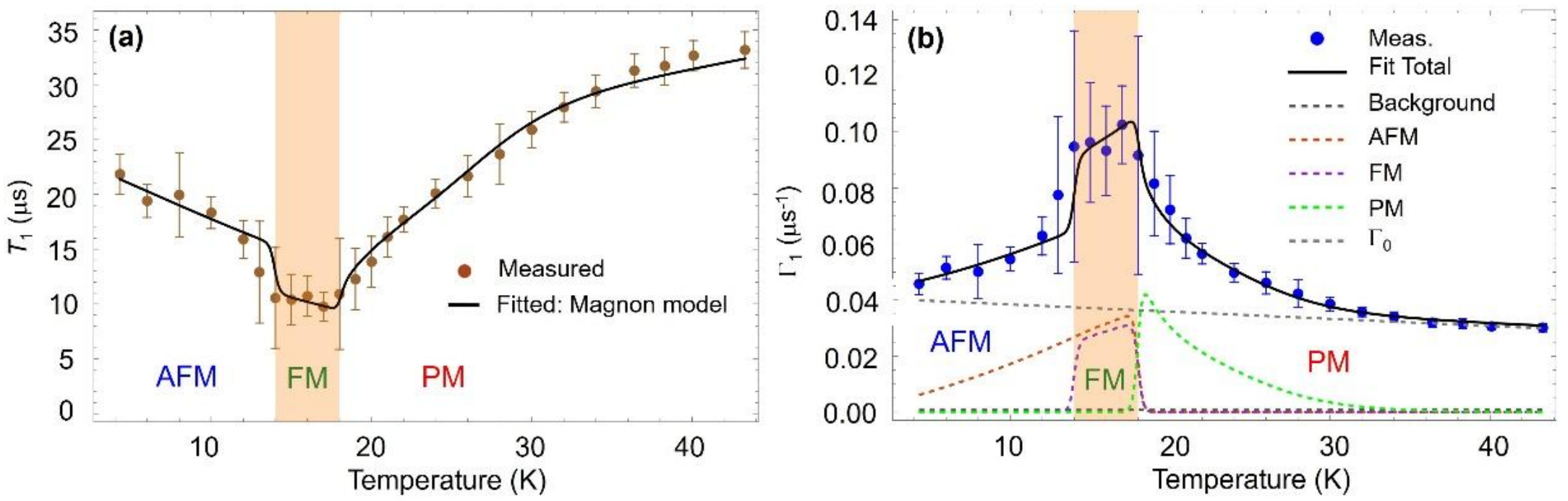

**Figure 5.** (a) Measured (scattered) and fitted (solid line) NV $T_1$ vs temperature. The fitting was done through the weighted three-window magnon model at a magnetic field of 1.45 mT. (b) Measured (scattered) and fitted (solid line) $\Gamma_1$ vs temperature at a magnetic field of 1.54 mT. The dashed lines show the $\Gamma_1$ fit decomposition using the three (AFM, FM, PM) window magnon model.

Importantly, all contributions share a single temperature-dependent background, which strongly constrains the fit and prevents independent arbitrary offsets of individual channels. The crossover functions serve to isolate the effective fitting of different contributions, by coupling only through the common background. This avoids double counting and suppresses unphysical extrapolations of asymptotic forms. The extracted parameters should therefore be regarded as effective quantities rather than directly interpretable physical constants. In this way, the decomposition does not simply separate the data into fitting terms but resolves how qualitatively different fluctuation processes contribute to the NV relaxation at the $CrCl_3$ magnetic phase transitions. This phenomenological decomposition therefore supports a marked change in the dominant relaxation physics across the crossover, consistent with FMR measurements, where the same temperature range is accompanied by clear changes in both the linewidth and the resonance field.

## Conclusions

To conclude, we investigated the spin dynamics of $CrCl_3$ nanoflakes across antiferromagnetic, ferromagnetic, and paramagnetic phases using cryogenic NV-center–based magnetometry and broadband FMR spectroscopy. By combining ODMR, Rabi measurements, and $T_1$ relaxometry, we access magnetic fluctuations in GHz frequencies and directly track their evolution across the closely spaced magnetic phase transitions of $CrCl_3$. A central result of this study is the marked enhancement of magnetic noise in the FM regime in comparison to AFM and PM phases, evidenced by the reduction of ODMR contrast, collapse of the coherent Rabi oscillations, and a strong increase of the NV relaxation rate $\Gamma_1$. Temperature-dependent FMR spectroscopy of $CrCl_3$ microcrystals shows 4-15 GHz spin dynamics with an increased linewidth up to 24 mT ($\alpha$ of $\sim 10^{-3}$) in the FM regime, confirming the presence of dense magnon bath that leads to increased NV $\Gamma_1$.

We used a phenomenological model to describe the spin-noise-induced NV relaxation, where the antiferromagnetic phase is represented by a gapped magnon dispersion, the ferromagnetic phase by a quadratic Kittel-like magnon dispersion, and the paramagnetic phase by a Curie–Weiss-like susceptibility term. The fitted $\Gamma_1(T)$ reproduces the measured NV relaxation well and supports a crossover in the dominant relaxation mechanism across the magnetic phase diagram.

These results establish $CrCl_3$ as a model platform for studying magnetic phase-dependent magnetic noise in vdW magnets and provide key input for evaluating its suitability in magnonic and hybrid quantum–magnon devices.[16,20–22,67,68] Beyond identifying phase-dependent signatures, this work highlights the capability of NV magnetometry to probe non-coherent spin dynamics and magnetic noise in layered vdW magnets, the study of field- and strain-tuned phase diagrams, and the investigation of fluctuation-driven phenomena beyond long-range order.[36–38,69]

## Methods

**Sample preparation.** Single crystals of $CrCl_3$ were grown using a self-transport technique starting from commercially available $CrCl_3$ powder (Thermo Fisher, 99.9%). Approximately 0.6 g of $CrCl_3$ was loaded into a 20 cm long quartz ampoule. The ampoule was sealed under vacuum and placed horizontally inside a tube furnace such that the starting material was positioned near the center (hot zone) of the furnace, while the opposite end of the tube extended toward the cooler region near the furnace opening. The furnace was heated to 973 K over 24 h and kept at this temperature for an extra 24 hours to guarantee homogenization. Successively, a temperature gradient of 723 – 823 K (hot-to-cold end) was established and maintained for several days to facilitate crystal growth via vapor transport. Large violet-colored $CrCl_3$ single crystals formed at the cooler end of the tube, see the inset of Figure S1.1 in the Supporting Information.[11] To obtain larger crystals, the growth procedure was repeated under identical conditions. Millimeter-sized crystals of $CrCl_3$ were collected and stored in an Ar-filled glovebox.

**X-ray Diffraction.** Single-crystal X-ray diffraction (XRD) spectroscopy was done at room temperature on $CrCl_3$ crystals, see Supporting Information Section S1. $CrCl_3$ crystallizes in the monoclinic crystal unit cell with space group C2/m. The refined lattice parameters are $a$ = 5.9672(8) Å, $b$ = 10.3322(10) Å, $c$ = 6.1311(9) Å, $\alpha$ = 90°, $\beta$ = 108.569(15)°, and $\gamma$ = 90°, yielding a unit cell volume of 358.33(8) Å$^3$. These values agree with previously measured values,[63,70] confirming the phase purity and structural integrity of the synthesized crystals.

**FMR spectroscopy of $CrCl_3$ crystals.** The spin dynamics of the $CrCl_3$ is studied using ferromagnetic resonance (FMR) spectroscopy[34] in a closed-cycle cryostat. The $CrCl_3$ crystal was mounted on a coplanar waveguide (CPW) inside a glove box to prevent degradation and secured with a mechanical clamp, ensuring environmental protection and good thermal contact with the CPW's copper surface. A temperature sensor was attached next to the sample for precise temperature monitoring. The CPW was connected to a microwave source via coaxial cables, and the transmitted signal was detected using a broadband microwave detector. For the FMR detection, we fixed the MW frequency and swept the applied magnetic field $H$. A Lock-in detection scheme was used to improve the signal-to-noise ratio.[34] We repeat the magnetic resonance measurements at different temperatures, across the FM-PM phase transition.

## Acknowledgements

K.A., J.W., and A.L. acknowledge the National Science Foundation (NSF) through Award 2328822. A.L. thanks for the additional support from NSF Award 2521415. T.L. and X.H. acknowledge the support of NSF Award No. DMR-2118828. A.L. and X.H. acknowledge the support of University of Nebraska-Lincoln (UNL) Grand Challenges catalyst award entitled "Quantum Approaches addressing Global Threats". The research done at UNL was performed in part in the Nebraska Nanoscale Facility: National Nanotechnology Coordinated Infrastructure and the Nebraska Center for Materials and Nanoscience (and/or NERCF), supported by NSF Award 2025298. I.F. acknowledges funding from the European Regional Development Fund, project No. 1.1.1.3/1/24/A/166.

## Conflict of Interest

The authors declare no conflict of interest.

## Author Contributions

A.L. and J.K. conceived the concept, designed the experiments, and supervised the project. B.H. and J.K. performed the NV measurements. S.I. fabricated the MW loops and assisted in NV measurements. I.F. performed the analysis of the NV relaxometry data and the magnetic noise modeling. K.P. and J.W. synthesized the $CrCl_3$ crystals and performed XRD measurements. P.B.K., A.E.A., and K.A. performed FMR measurements on $CrCl_3$ crystals. T.L., A.P., and X.H. exfoliated the $CrCl_3$ flakes and performed optical and topography characterization. A.L. and J.K. wrote the manuscript with contributions and feedback from all authors.

## Data Availability Statement

The data that support the findings of this study are available from the corresponding author upon reasonable request.

## Keywords

chromium chloride, nitrogen vacancy, $T_1$ relaxometry, spin dynamics, ferromagnetic resonance spectroscopy.

# Supporting Information

## Section 1. Synthesis and characterization of $CrCl_3$ crystals and flakes

### S1.1 Powder X-ray diffraction of $CrCl_3$ crystals

Room temperature x-ray diffraction (XRD) spectra were measured using the Rigaku Mini Flex 6G diffractometer (Cu Kα radiation at a wavelength λ of 1.5406 Å), as detailed in reference [1]. Because a flake-like $CrCl_3$ single crystal was used for the measurement, the diffraction pattern exhibits only (00l) reflections, indicating a strong preferred orientation along the c-axis. The observed peak positions agree well with the simulated pattern calculated from the reported crystallographic data, confirming phase purity and consistency with the reported structure.

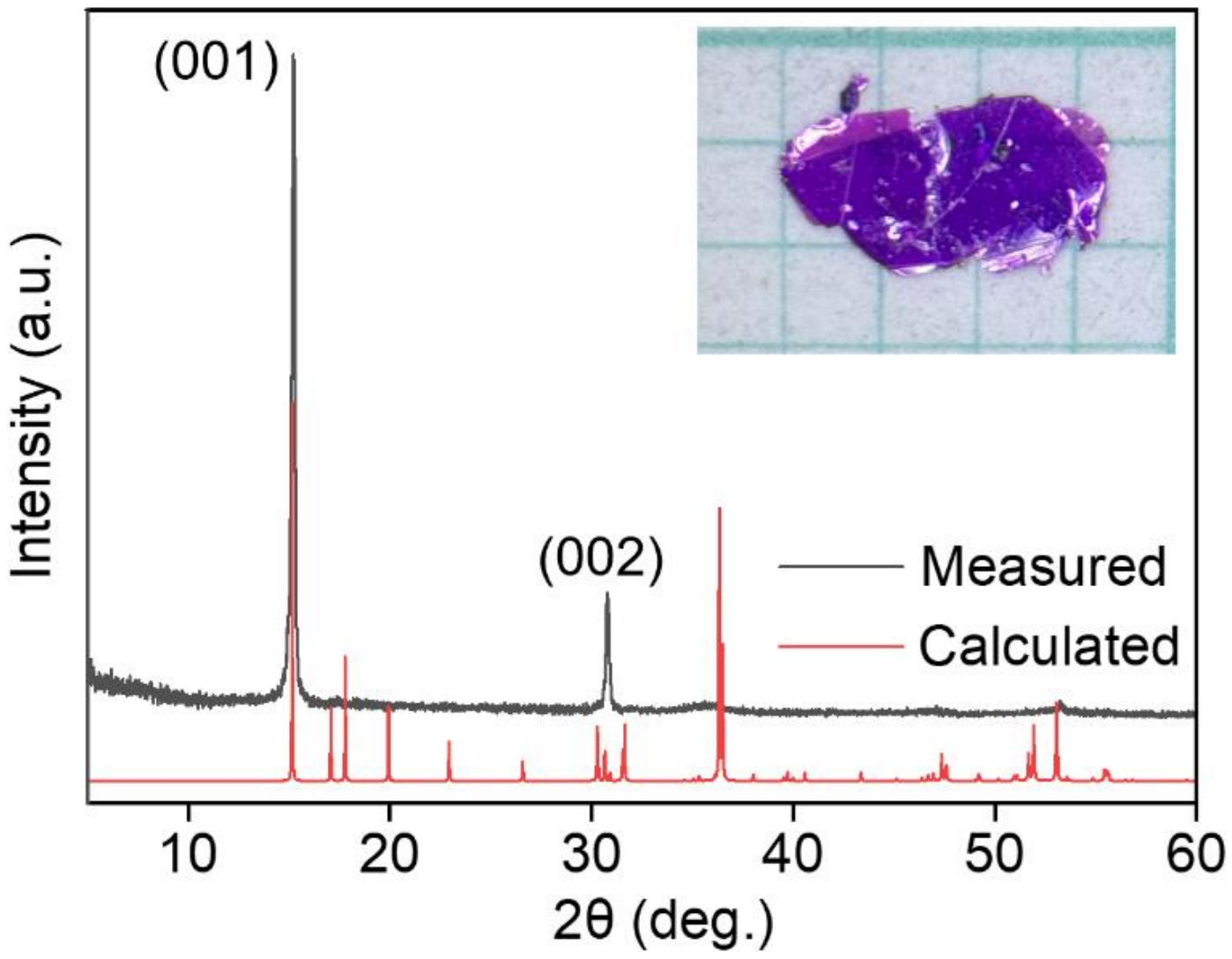

**Figure S1.1:** Powder X-ray diffraction pattern of $CrCl_3$ crystals. Inset: Optical image of a violet $CrCl_3$ single crystal.

### S1.2. Single Crystal X-ray Diffraction (SCXRD)

Room temperature single-crystal XRD spectra were performed on $CrCl_3$ using the Rigaku XtaLAB mini II diffractometer (Mo $K_{\alpha}$ radiation at a wavelength λ of 071073 Å). We performed data reduction by using the CrysAlisPro software,[2] and crystal-structure refinement by using the SHELX[3] suite of programs within the Olex2 v1.5 interface.[4] See Table S1 for more details.

**Table S1**: Lattice parameters of $CrCl_3$ crystal obtained from SCXRD

| Compound | Experimental | Theoretical |
|---|---|---|
| Empirical Formula | $CrCl_3$ | $CrCl_3$[5,6] |
| Formula weight | 184.35 | - |
| Temperature | 293 K | 293 K |
| Radiation wavelength | Mo $K_{\alpha}$, 0.71073 Å | Mo $K_{\alpha}$, 0.70926 Å |

| Crystal system | monoclinic | monoclinic |
|---|---|---|
| Space group | C2/m | C2/m |
| a (Å) | 5.9672(8) | 5.9617 |
| b (Å) | 10.3322(10) | 10.3361 |
| c (Å) | 6.1311(9) | 6.1244 |
| α (°) | 90 | 90 |
| β (°) | 108.569(15) | 108.6 |
| γ (°) | 90 | 90 |
| V ($Å^3$) | 358.33(8) | 357.7 |
| Z | 4 | 4 |
| $D_c$ (g $cm^{-3}$) | 3.417 | - |
| μ ($mm^{-1}$) | 6.591 | - |
| $R_1$,<br>$wR_2$ ($I > 2\sigma(I)$) | 0.1064<br>0.2385 | - |
| $R_1$,<br>$wR_2$ (all data) | 0.1176<br>0.2559 | - |

$R_1 = \sum||F_o| - |F_c||/\sum|F_o|$; $wR_2 = [\sum[w(F_o^2 - F_c^2)^2]/\sum[w(F_o^2)^2]]^{1/2}$, and $w = 1/[\sigma^2 F_o^2 + (A\cdot P)^2 + B\cdot P]$, $P = (F_o^2 + 2F_c^2)/3$, where A and B are weight coefficients

**S1.3 Exfoliation and transfer of $CrCl_3$ flakes**

Bulk $CrCl_3$ crystals were mechanically exfoliated using scotch tape and transferred on top of a silicone gel film (PF-40-X4), following the recipe used in reference [7]. Large flakes with thickness below 65 nm were identified by optical microscopy based on optical contrast.[8,9] The flake thickness was determined by atomic force microscopy. The selected $CrCl_3$ flakes were transferred onto a diamond substrate via gel film using the high-temperature dry transfer technique.[10,11] During transfer, the flake on the gel film was first aligned with the target location on the diamond substrate. After the gel film was in contact with the substrate, a soft pressure was applied, and the stage was warmed up to 333 K for 54 minutes to improve interfacial stickiness. The gel film was then slowly retracted, leaving the $CrCl_3$ flake on the diamond substrate. Figure S1.3a shows the topography image studied with nitrogen vacancy (NV) microscope in the main text (Figure 1). The corresponding transverse cut of the flake indicates a thickness of 60 ± 1 nm. We also performed NV measurements on another $CrCl_3$ flake, discussed below in Section S4. Figures S1.4a and S1.4b show the optical and AFM images of a selected $CrCl_3$ flake studied by nitrogen-vacancy (NV) microscopy (see Section S4). A transverse cut of the AFM image in Figure S1.4c shows a thickness of 42 ± 0.5 nm.

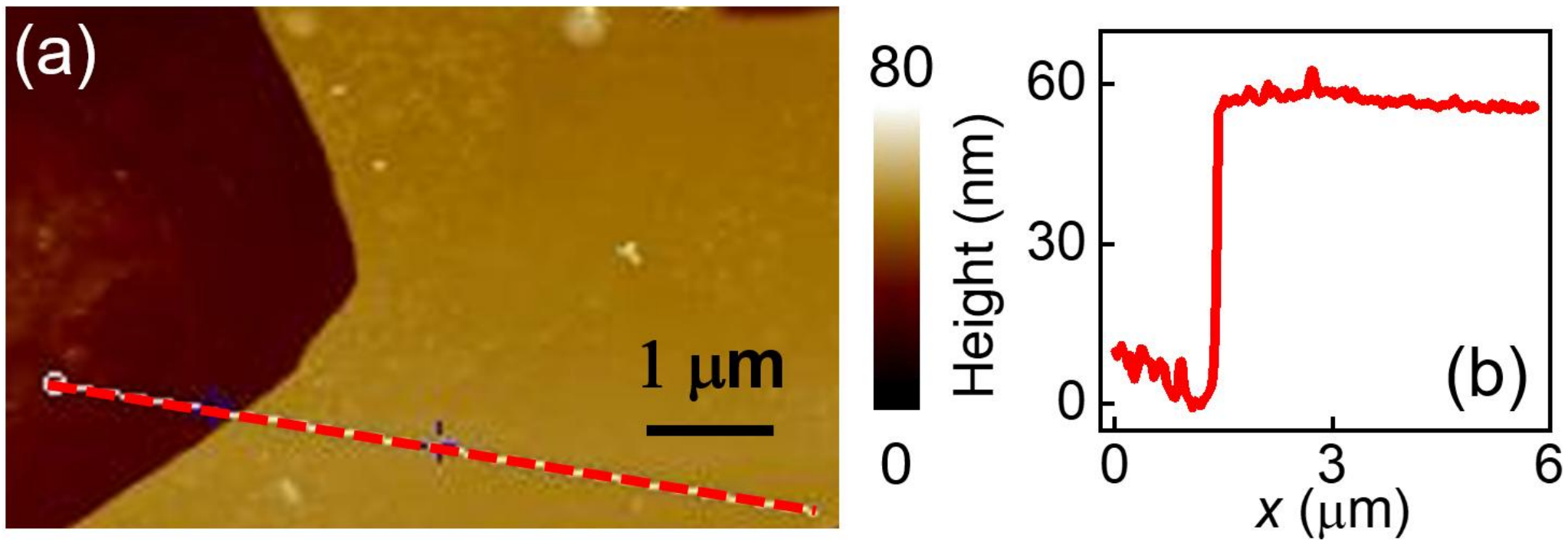

**Figure S1.3:** (a) Topography image of the $CrCl_3$ flake studied in the main text (Figure 1). (b) Height profile of the flake in (a) along the dashed red line showing a height of 60 ± 1 nm.

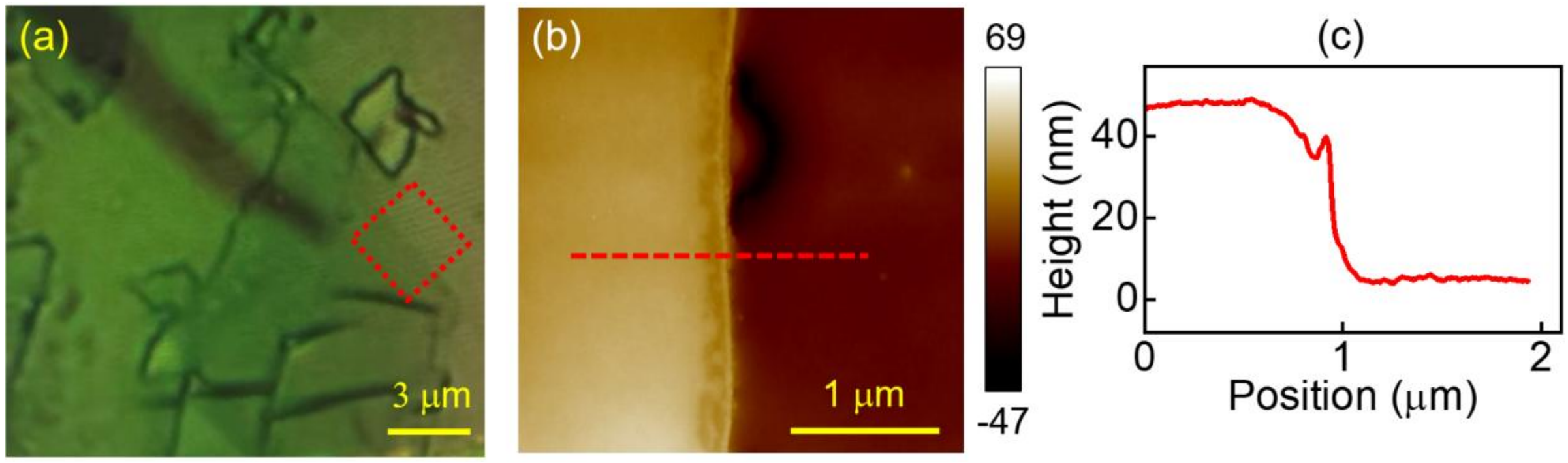

**Figure S1.4:** Optical (a) and topography (b) images of a selected $CrCl_3$ flake studied with NV microscopy. (c) Height profile of the flake in (b) along the dashed red line showing a height of 42 ± 0.5 nm.

## S2. Autofluorescence from $CrCl_3$ flakes

While performing NV optically detected magnetic resonance (ODMR) measurements on the $CrCl_3$ flakes, we found a considerable autofluorescence coming from the flakes excited at a laser power of 36 mW (beam size ~23 μm), see Figure S2a. To check the spectral properties of this autofluorescence, we used a confocal fluorescence microscope[8] to measure the fluorescence intensity on a selected $CrCl_3$ flake excited with a green laser (532 nm) at a laser power of 1 mW. A homogeneous fluorescence was obtained across the flake (Figure S2b) with a high intensity at wavelengths above 720 nm (Figure S2c). A detailed study of the optical properties of $CrCl_3$ bulk crystals showed a fluorescence in the wavelength range of 750 – 990 nm excited with a 532 nm laser, explained by bound ligand field excitons.[12] To completely cut the fluorescence from $CrCl_3$ flakes, we used a bandpass filter in the wavelength range of 650 – 710 nm. Figure S2d shows the widefield image of the same $CrCl_3$ flake (Figure S2a) with a fluorescence coming from NVs.

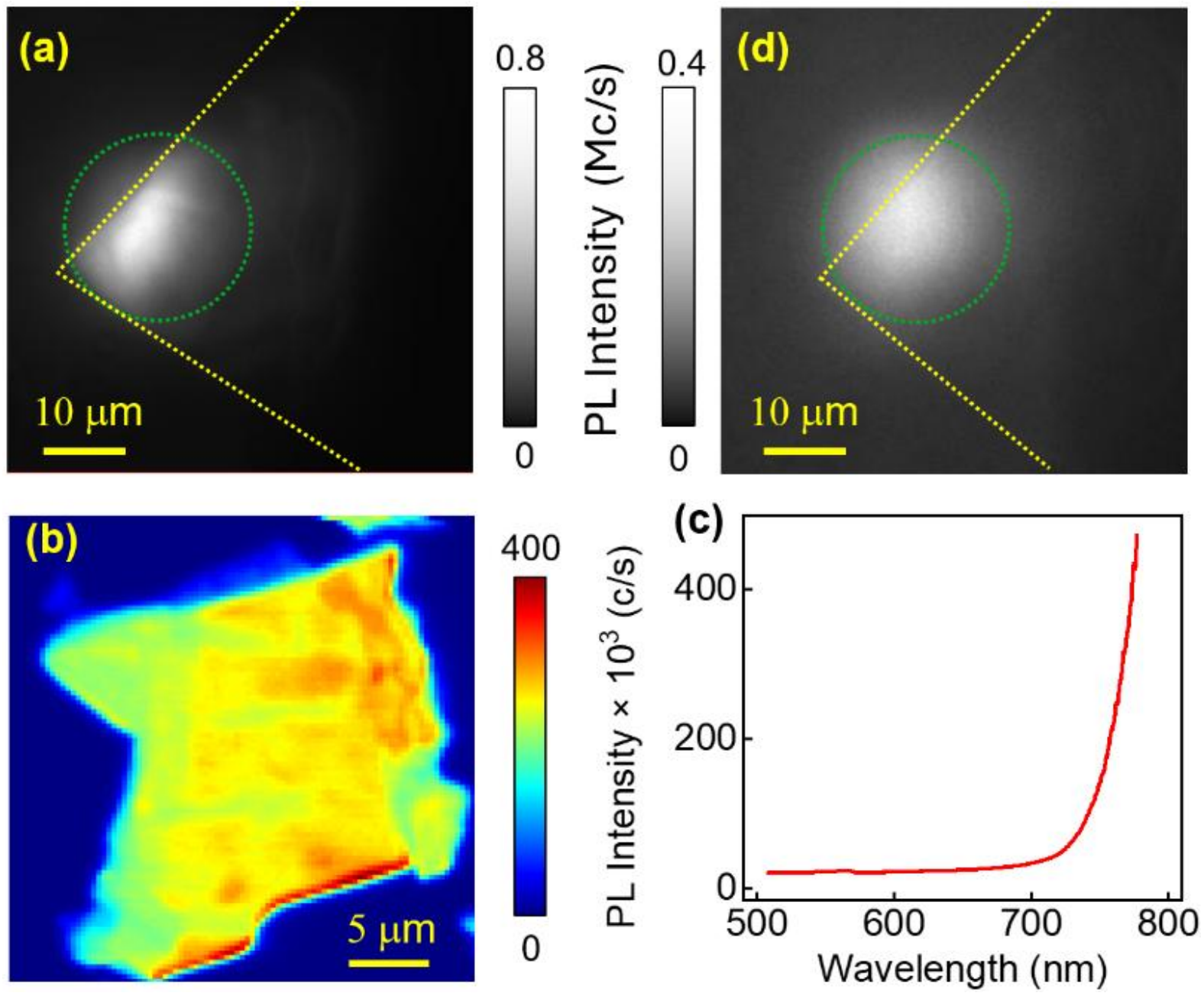

**Figure S2:** (a) Widefield fluorescence image of $CrCl_3$ flake (thickness of ~60 nm) studied in the main text excited at a 532 nm laser power of 36 mW. (b) Fluorescence image of a selected $CrCl_3$ flake showing a high autofluorescence. (c) Fluorescence spectrum (intensity vs wavelength) of the flake in (b). (d) Widefield fluorescence image of the same $CrCl_3$ flake in (a) after installing the bandpass filter in the wavelength range of 650 – 710 nm.

### S3. NV measurements on 60 nm thick $CrCl_3$ flake

To obtain the ODMR resonance frequencies, the normalized contrast, and full width at half maximum (FWHM), the measured ODMR spectra (Figure S3.1) were fitted using Lorentzian line shapes, see Figure S3.1. The fitting results corresponding to the $f_-$ peak are shown separately in Figure S3.2.

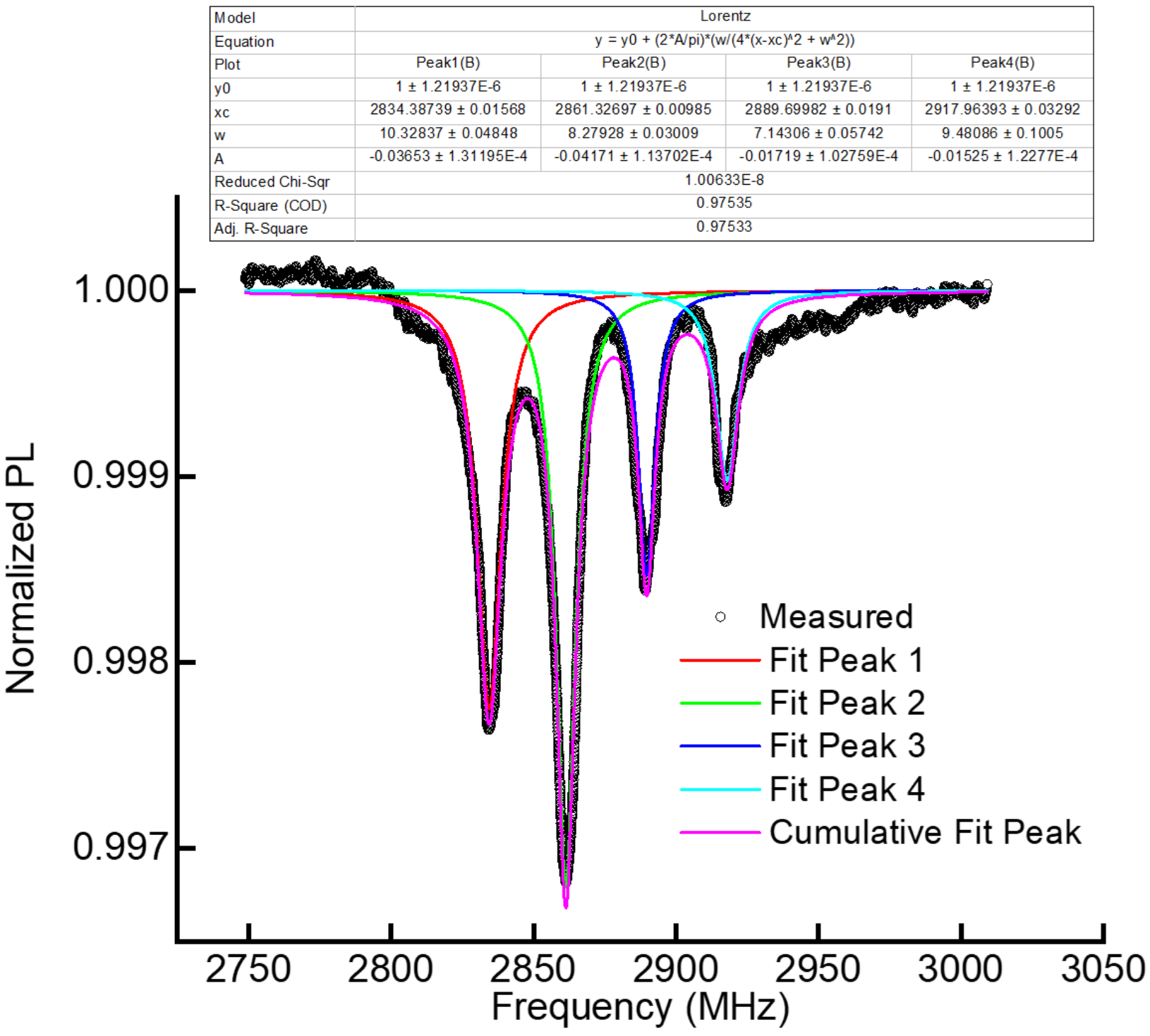

| Model | Lorentz | | | |
|---|---|---|---|---|
| Equation | y = y0 + (2*A/pi)*(w/(4*(x-xc)^2 + w^2)) | | | |
| Plot | Peak1(B) | Peak2(B) | Peak3(B) | Peak4(B) |
| y0 | 1 ± 1.21937E-6 | 1 ± 1.21937E-6 | 1 ± 1.21937E-6 | 1 ± 1.21937E-6 |
| xc | 2834.38739 ± 0.01568 | 2861.32697 ± 0.00985 | 2889.69982 ± 0.0191 | 2917.96393 ± 0.03292 |
| w | 10.32837 ± 0.04848 | 8.27928 ± 0.03009 | 7.14306 ± 0.05742 | 9.48086 ± 0.1005 |
| A | -0.03653 ± 1.31195E-4 | -0.04171 ± 1.13702E-4 | -0.01719 ± 1.02759E-4 | -0.01525 ± 1.2277E-4 |
| Reduced Chi-Sqr | 1.00633E-8 | | | |
| R-Square (COD) | 0.97535 | | | |
| Adj. R-Square | 0.97533 | | | |

**Figure S3.1:** Measured and fitted NV ODMR peaks at a magnetic field of 1.45 mT and at a temperature of 28 K on the 60 nm $CrCl_3$ flake.

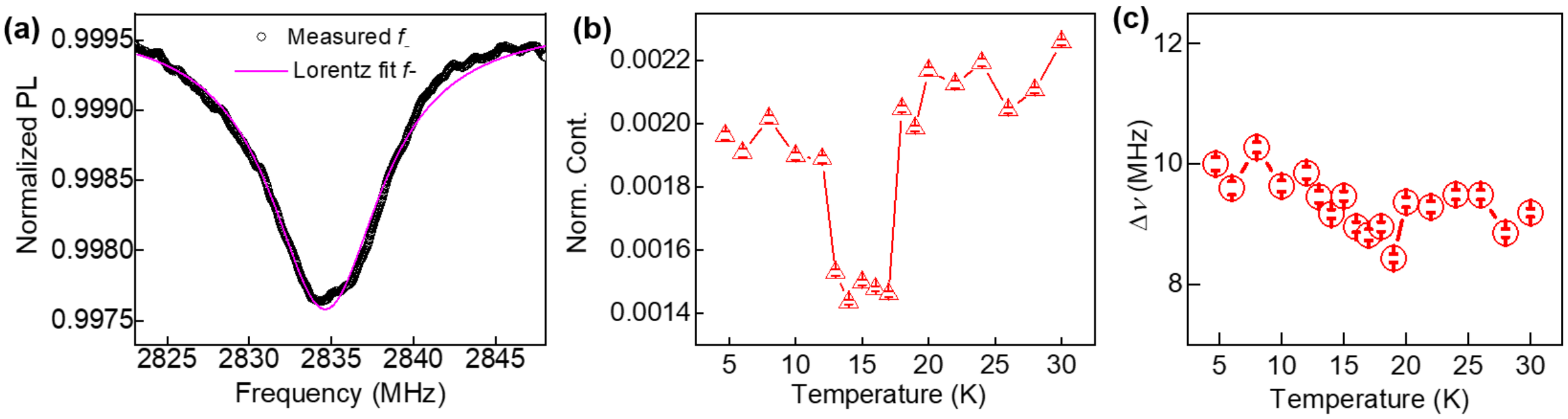

**Figure S3.2:** (a) Measured and fitted NV ODMR $f_-$ peak at a magnetic field of 1.45 mT and at 28 K. ODMR contrast (b) and FWHM (c) vs temperature, obtained by fitting the NV $f_-$ peak.

The NV Rabi and $T_1$ measurements vs temperature on the 60 nm thick $CrCl_3$ flake, studied in the main text, are plotted below at magnetic fields ranging from 1.45 to 40.38 mT on the flake, and at 1.45 mT off the flake as control data.

***NV Rabi and $T_1$ measurements at 1.45 mT off the $CrCl_3$ flake***

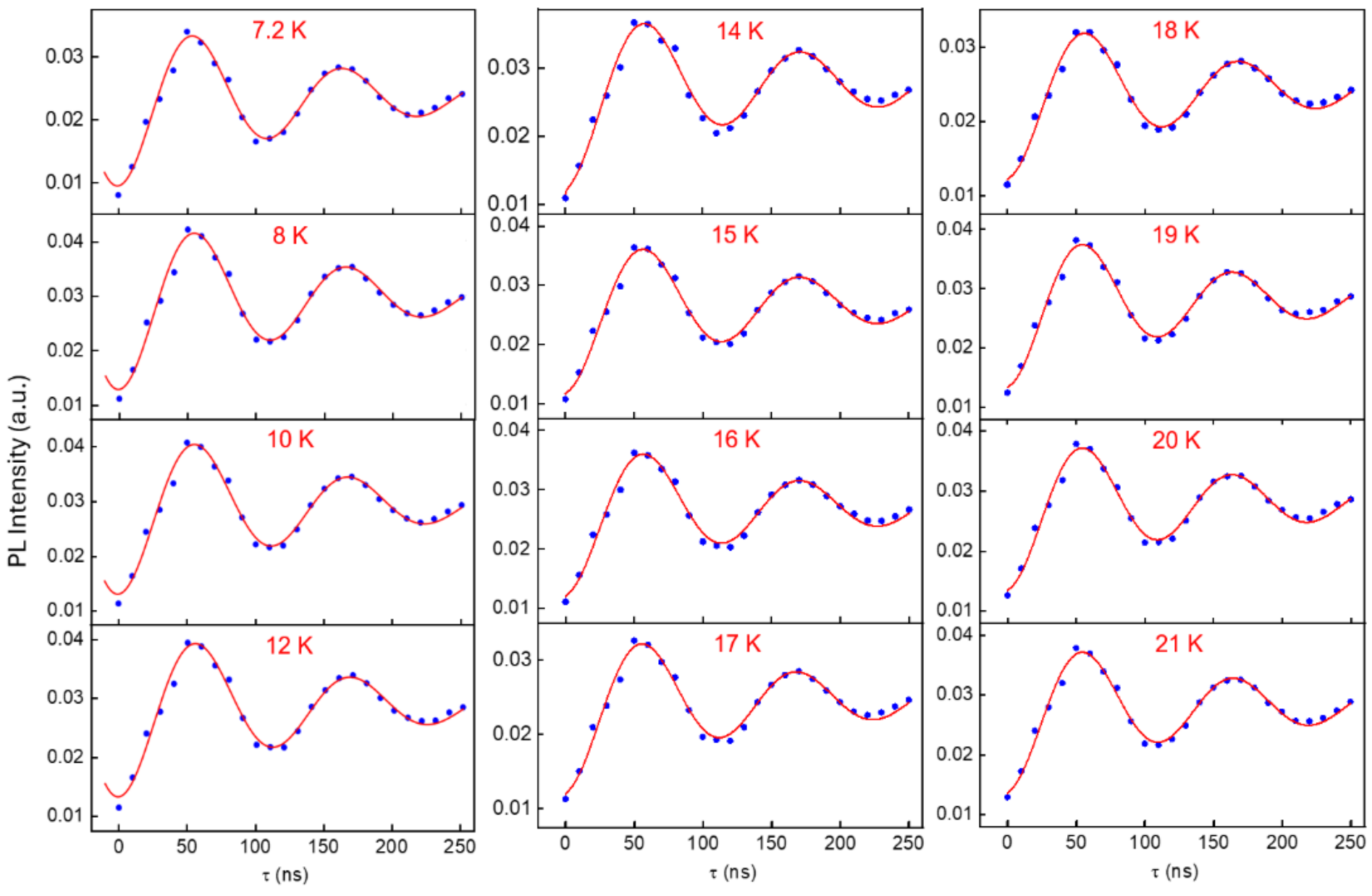

**Figure S3.1.1:** Rabi curves off the $CrCl_3$ flake vs temperature (7.2 – 21 K) at 1.45 mT.

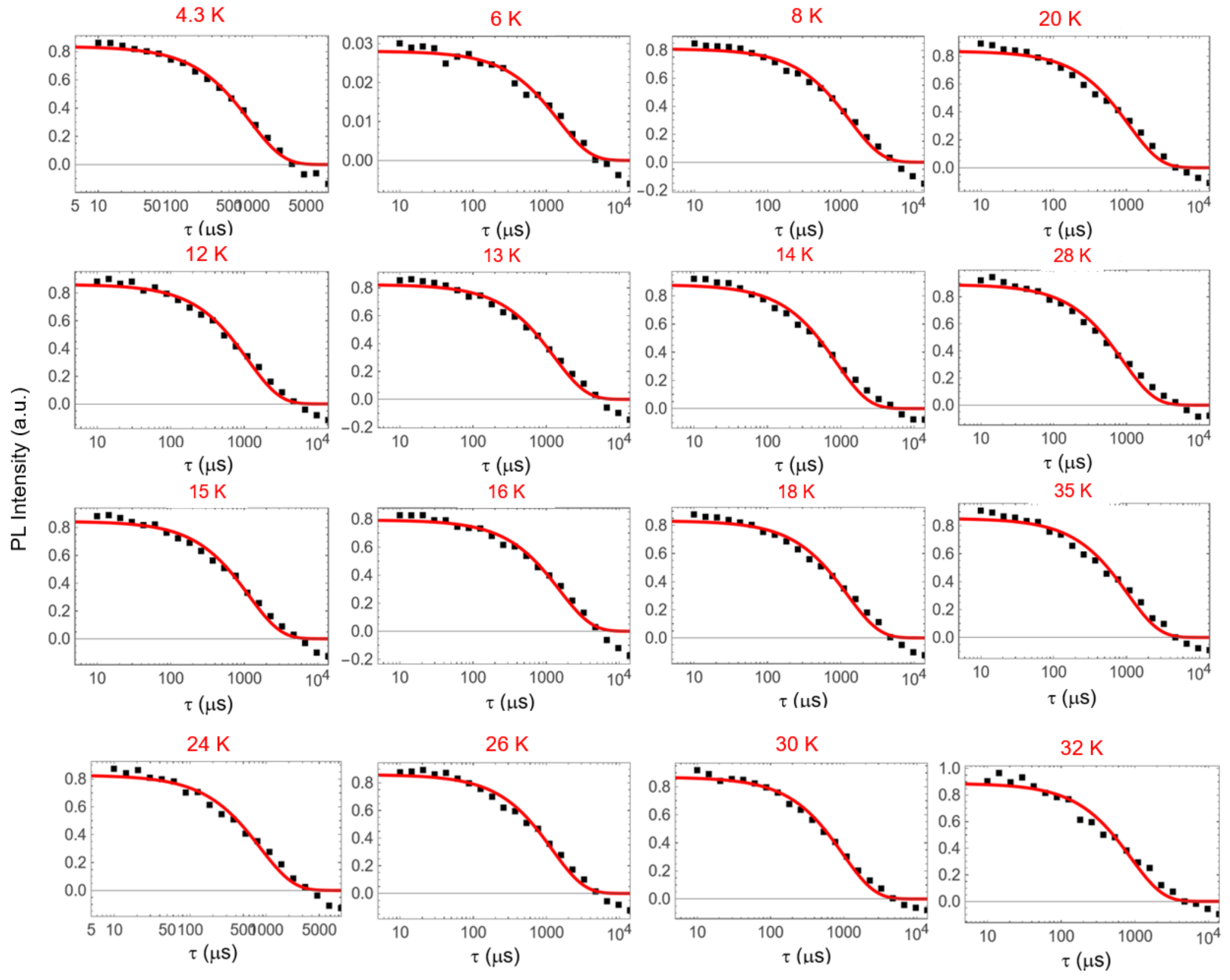

**Figure S3.1.2:** $T_1$ curves off the $CrCl_3$ flake vs temperature (4.3 – 32 K) at 1.45 mT.

### *NV Rabi and $T_1$ measurements at 1.45 mT on the $CrCl_3$ flake*

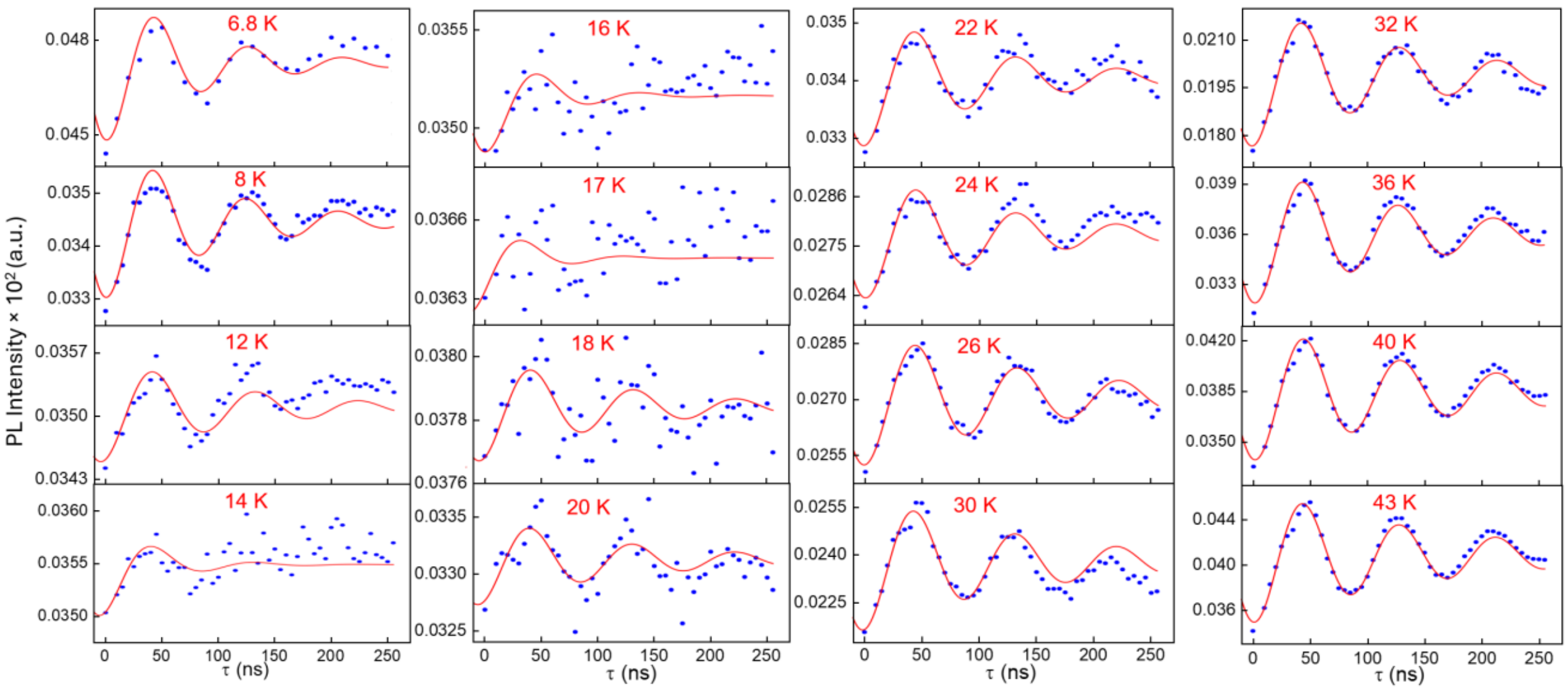

**Figure S3.2.1:** Rabi curves on the $CrCl_3$ flake vs temperature (6.8 – 43 K) at 1.45 mT.

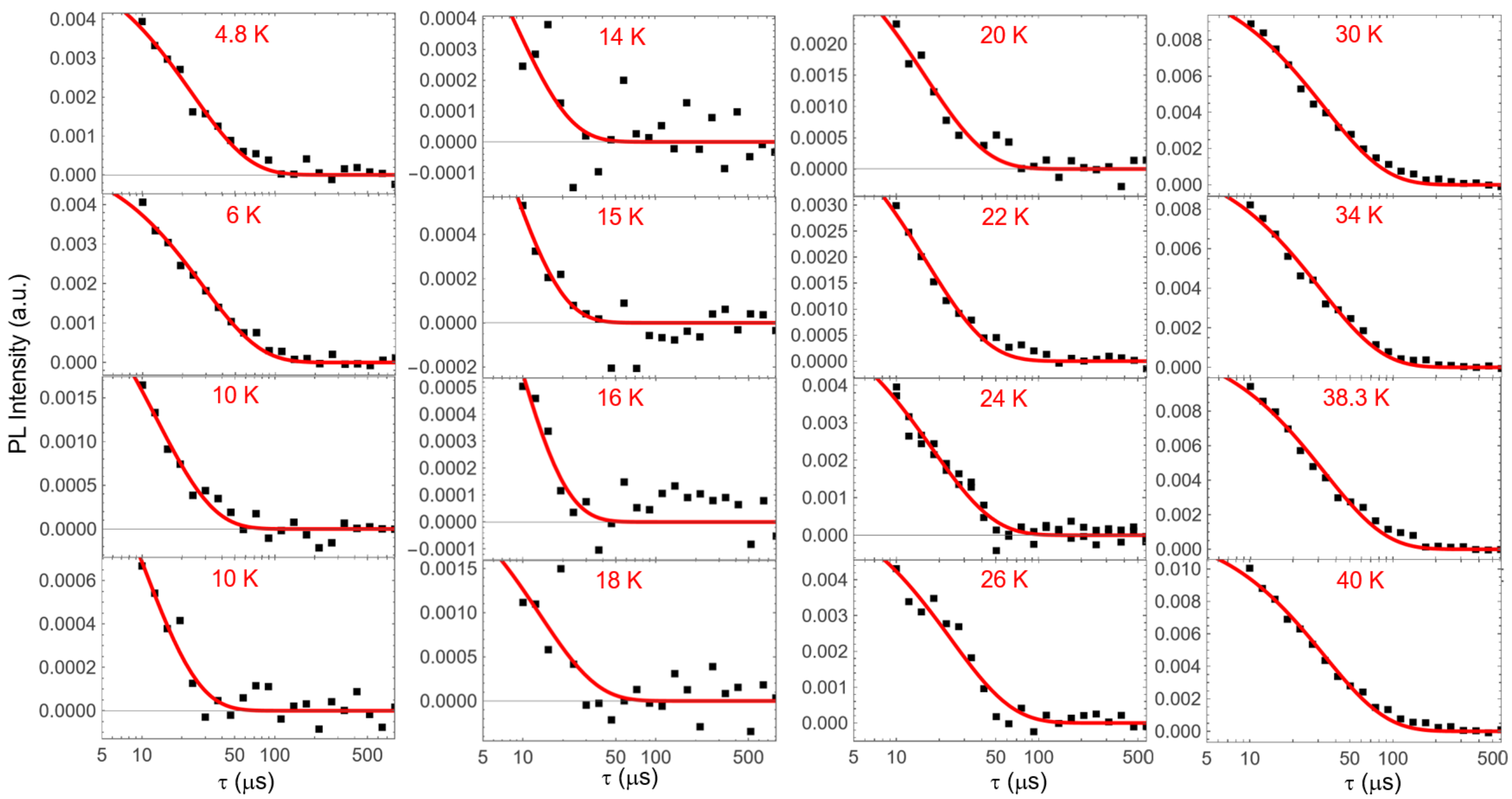

**Figure S3.2.2:** $T_1$ curves vs temperature (4.8 – 40 K) on the $CrCl_3$ flake at 1.45 mT.

***NV Rabi at 12.56 mT and $T_1$ measurements at 13.95 mT on the $CrCl_3$ flake***

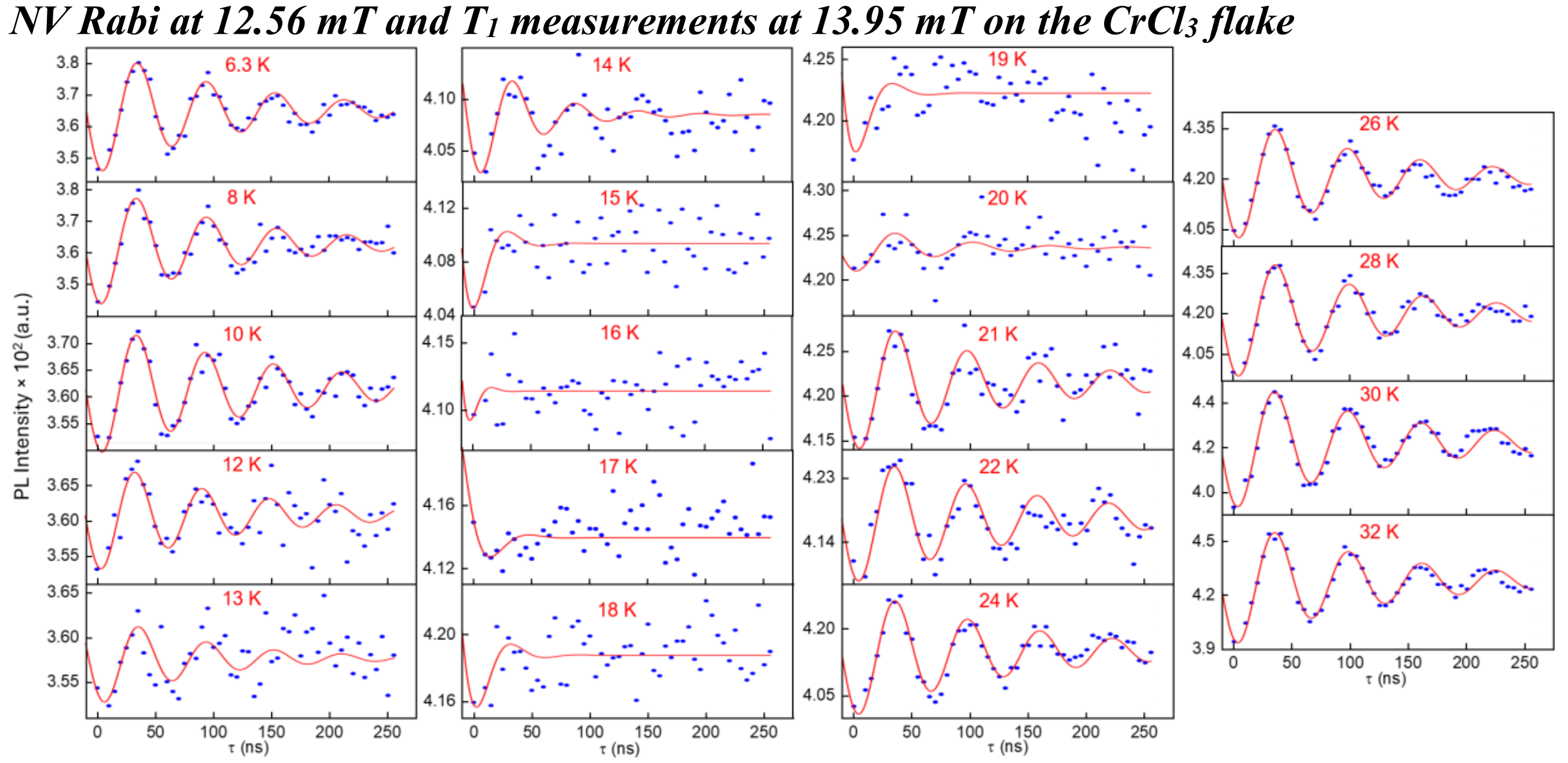

**Figure S3.3.1:** Rabi curves on the $CrCl_3$ flake vs temperature (6.3 – 32 K) at 12.56 mT.

We plot in Figure S3.3.2 both the Rabi frequency and Rabi field $B_1$ versus temperature at a magnetic field of 12.56 mT. There is no change in the $B_1$ amplitude in the ferromagnetic (FM) regime suggesting that the observed effects on Rabi oscillations are related to the increased decoherence from the ferromagnetic spin noise generated in the $CrCl_3$ flake, as discussed in the main text.

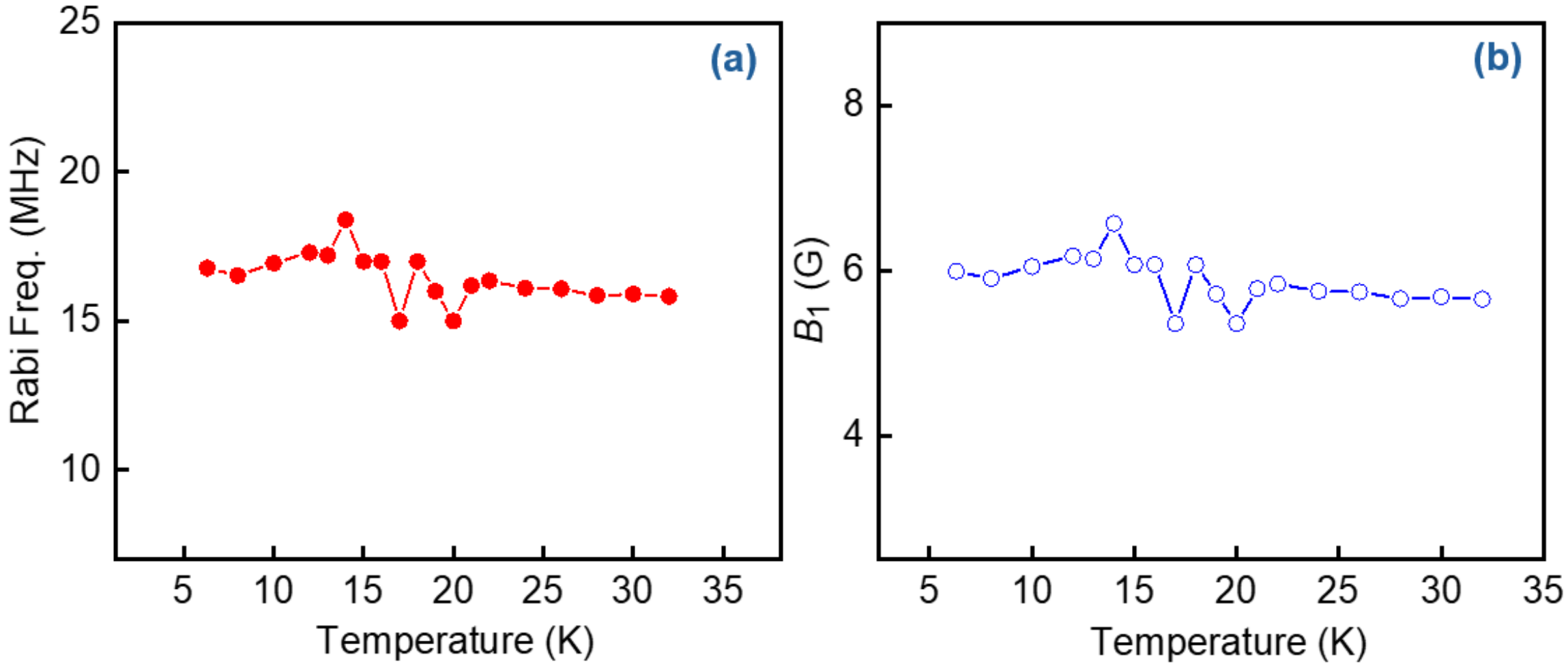

**Figure S3.3.2:** Rabi frequency (a) and $B_1$ field (b) and vs temperature at 12.56 mT.

Due to the slight bump of the permanent magnet outside the cryostat chamber after Rabi measurements and due to the long time required for $T_1$ measurements, we opted to realign the field close to the previous Rabi measurements and performed $T_1$ at the new field aligned at 13.95 mT. Figure S3.3.3 shows the corresponding $T_1$ curves at different temperatures.

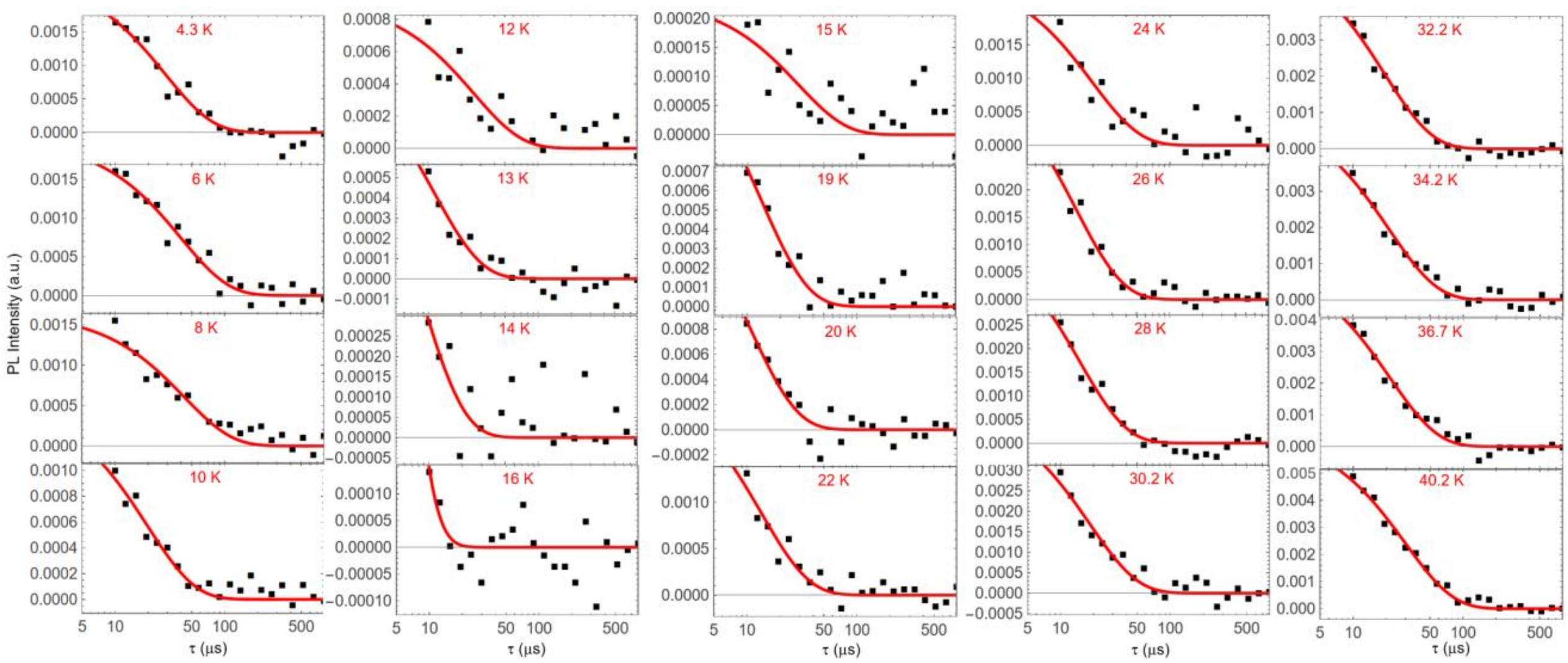

**Figure S3.3.3:** $T_1$ curves vs temperature (4.3 – 40.2 K) on the $CrCl_3$ flake at 13.95 mT.

### *NV Rabi and $T_1$ measurements at 24.51 mT on the $CrCl_3$ flake*

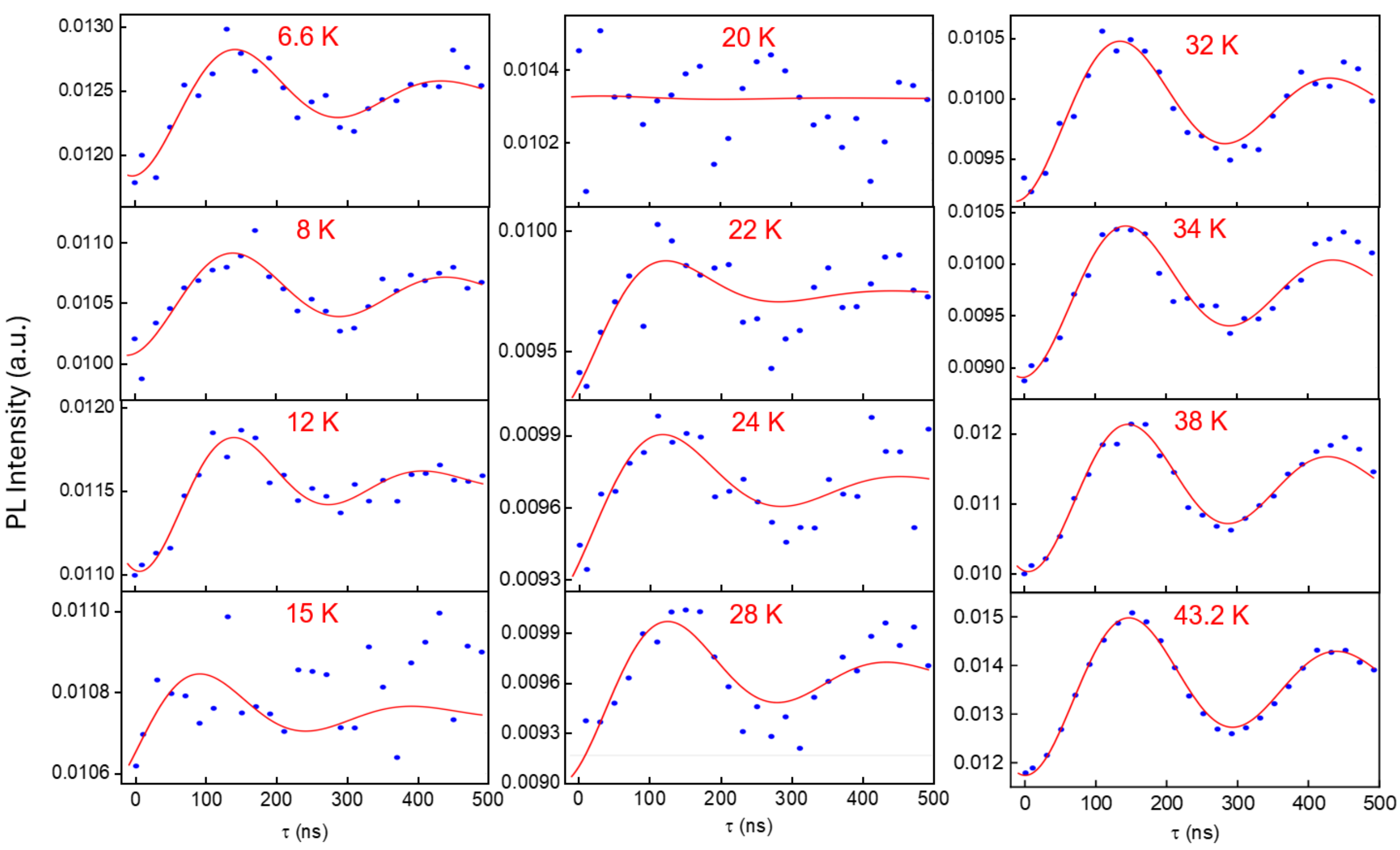

**Figure S3.4.1:** Rabi curves on the $CrCl_3$ flake vs temperature (6.6 – 43.2 K) at 24.51 mT.

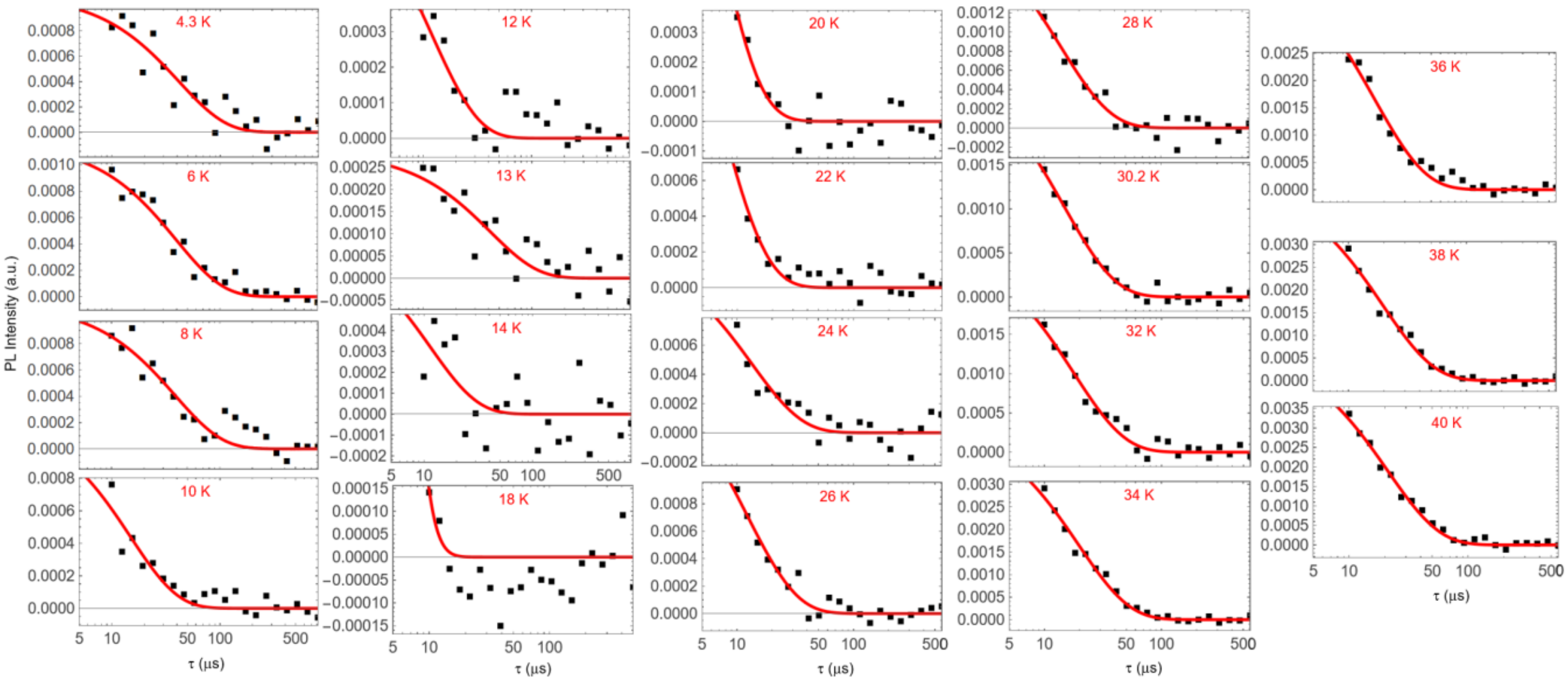

**Figure S3.4.2:** $T_1$ curves vs temperature (4.3 – 40.2 K) on the $CrCl_3$ flake at 24.51 mT.

### *NV Rabi and $T_1$ measurements at 40.38 mT on the $CrCl_3$ flake*

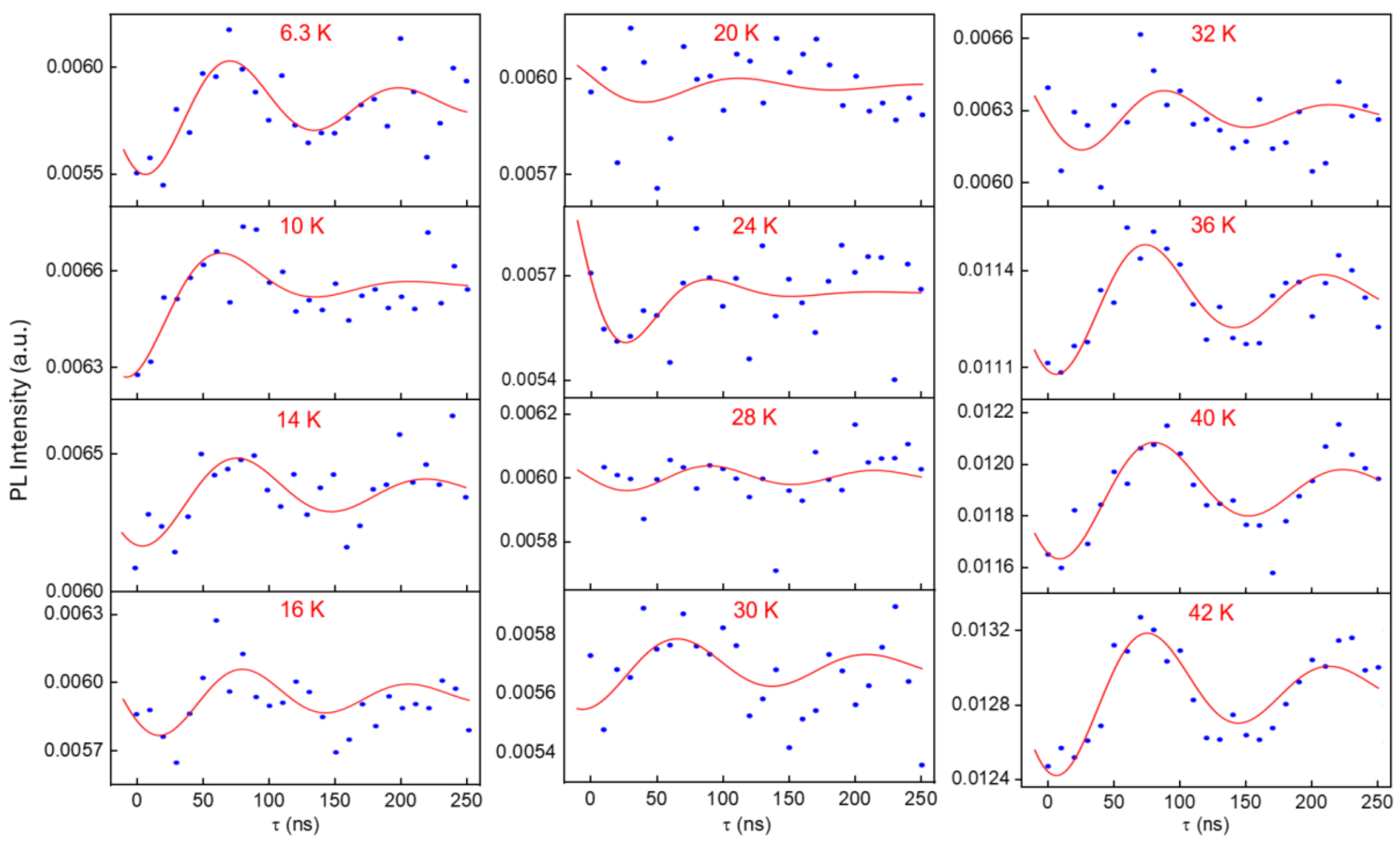

**Figure S3.5.1:** Rabi curves on the $CrCl_3$ flake vs temperature (6.3 – 42 K) at 40.38 mT.

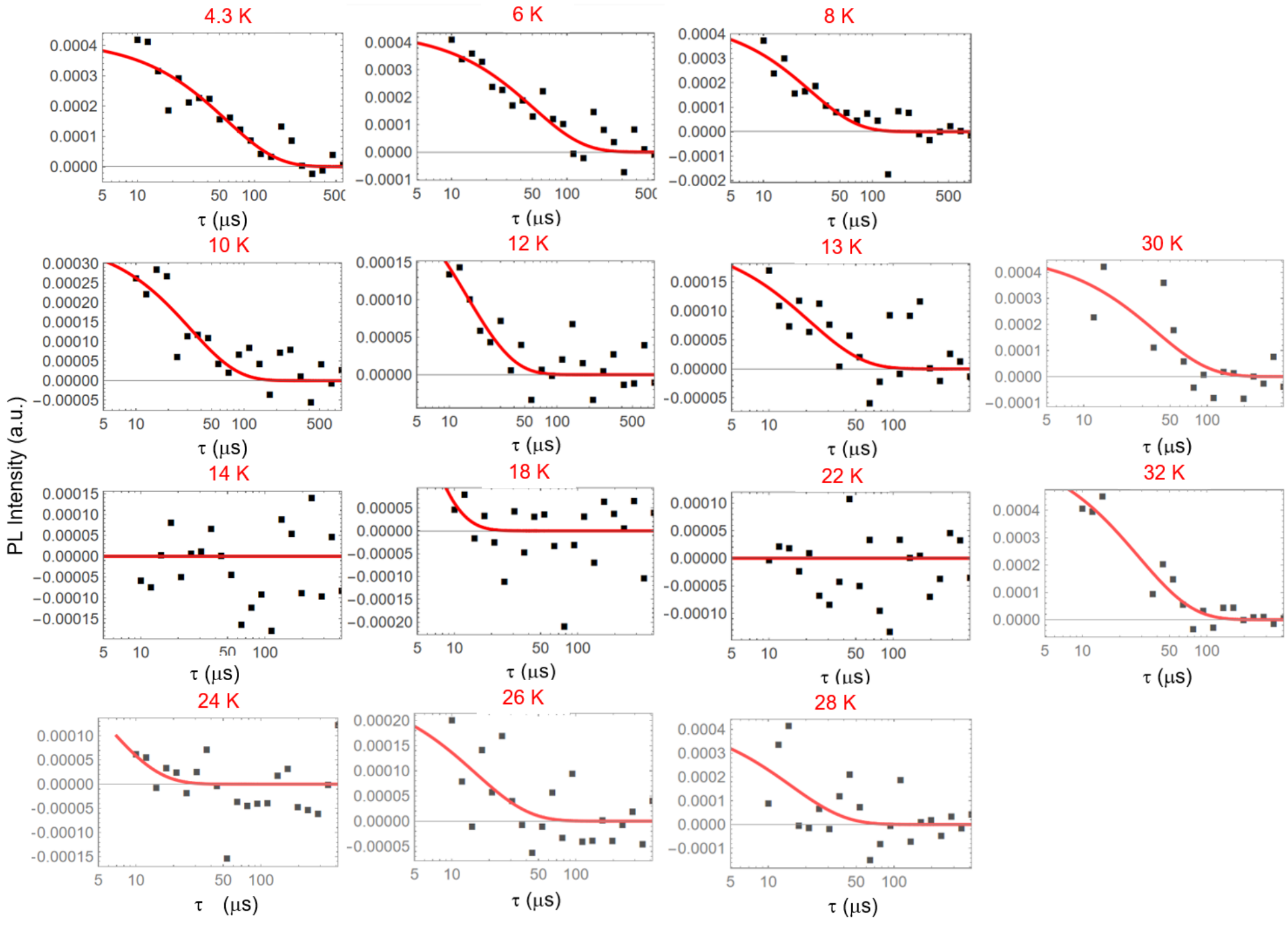

**Figure S3.5.2:** $T_1$ curves vs temperature (4.3 – 32 K) on the $CrCl_3$ flake at 40.38 mT.

Due to the high spin noise induced by the 60 nm $CrCl_3$ flake at higher magnetic field (> 20 mT), the Rabi amplitude, Rabi decay $T_{Rabi}$, and $T_1$ are plotted versus temperature separately from the low (< 20 mT) magnetic field measurements in Figures S3.6(a-c) at 24.51 mT and in Figures S3.6(d-f) at 40.38 mT. The observed broad (up to 24 mT) ferromagnetic resonance (FMR) response in the FM regime (see Figures 4c and 4d) indicates enhanced damping and large population of incoherent magnons in the $CrCl_3$ flake. Here, microwave (MW) excitation can generate strongly spatially fluctuating spin dynamics, leading to an increased magnetic noise with a spectral density at the NV location.[13,14]

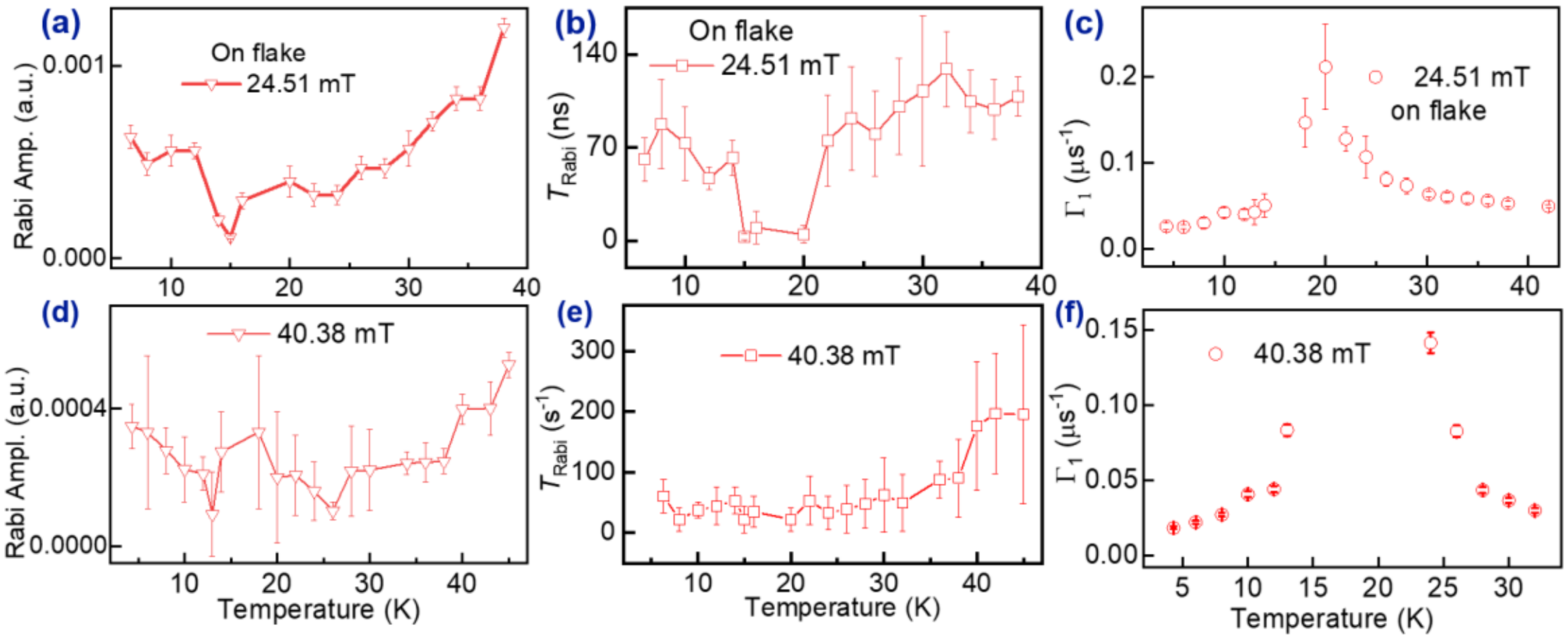

**Figure S3.6:** Rabi amplitude vs temperature at 24.51 mT (a) and 40.38 mT (d). Rabi decay vs temperature at 24.51 mT (b) and 40.38 mT (e). $\Gamma_1$ vs temperature at 24.51 mT (c) and 40.38 mT (f).

## S4. NV measurements on 42 nm thick $CrCl_3$ flake

We performed additional NV measurements on the 42 nm thick $CrCl_3$ flake at a magnetic field of 27.77 mT, shown in the optical and AFM measurements in Figure S1.4. Similar magnetic effects from the flake of the NV Rabi (amplitude, decay) and $T_1$ were obtained and discussed along the antiferromagnetic (AFM), FM, and paramagnetic (PM) transitions, studied in detail in the main manuscript. The effect of magnetic noise coming from the $CrCl_3$ flake seems to affect the NV Rabi in both AFM and FM, see Figure S4.1. However, it is observed only in the FM region, where $\Gamma_1$ rate increases by ~30 times near $T_C$ (Figure S4.2). This is explained by the spin waves excited by the MW in the FM region, see the main text and Supporting Information Section S5.

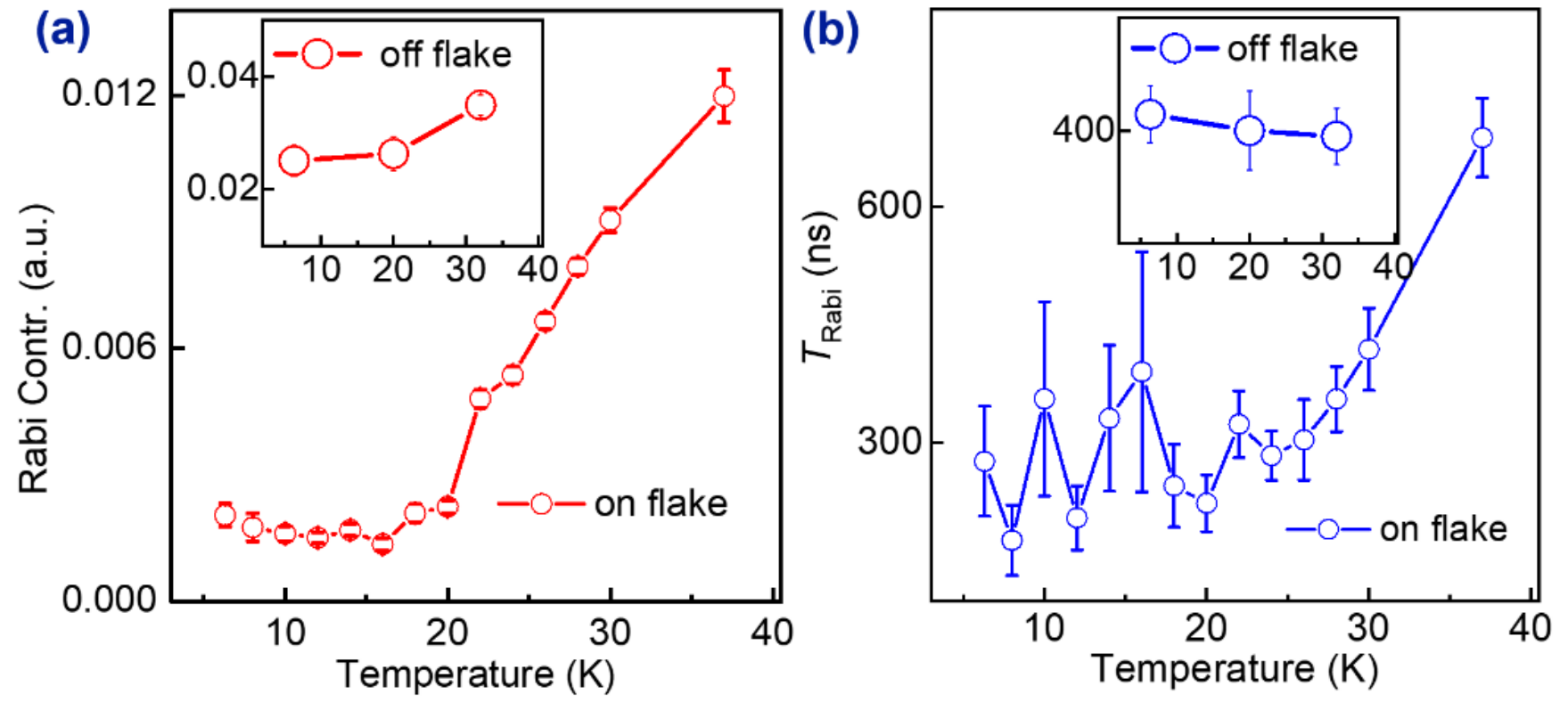

**Figure S4.1:** Rabi amplitude (a) and decay (b) on and off (inset) the $CrCl_3$ flake.

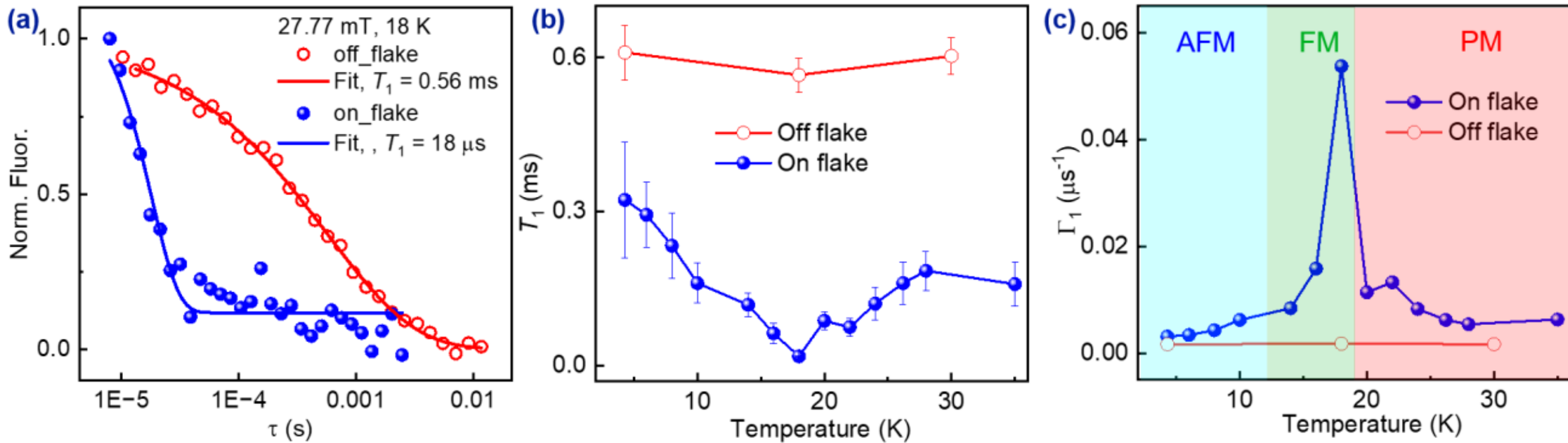

**Figure S4.2:** (a) $T_1$ curve off (open circles) and on (filled circles) the $CrCl_3$ flake at 27.77 mT and 18 K. $T_1$ (b) and $\Gamma_1$ (c) vs temperature off (open circles) and on (filled circles) the CrCl3 flake, confirming the AFM to FM to PM transitions.

The NV Rabi and $T_1$ measurements on the 42 nm thick $CrCl_3$ flake, discussed above, are plotted below at a magnetic field of 27.77 mT on and off the flake. A ~2 K heating was deduced and added to the Rabi curves due to the high power (10 W) MW excitation (see the main text for further details).

***NV Rabi and $T_1$ measurements at 27.77 mT off the 42 nm $CrCl_3$ flake***

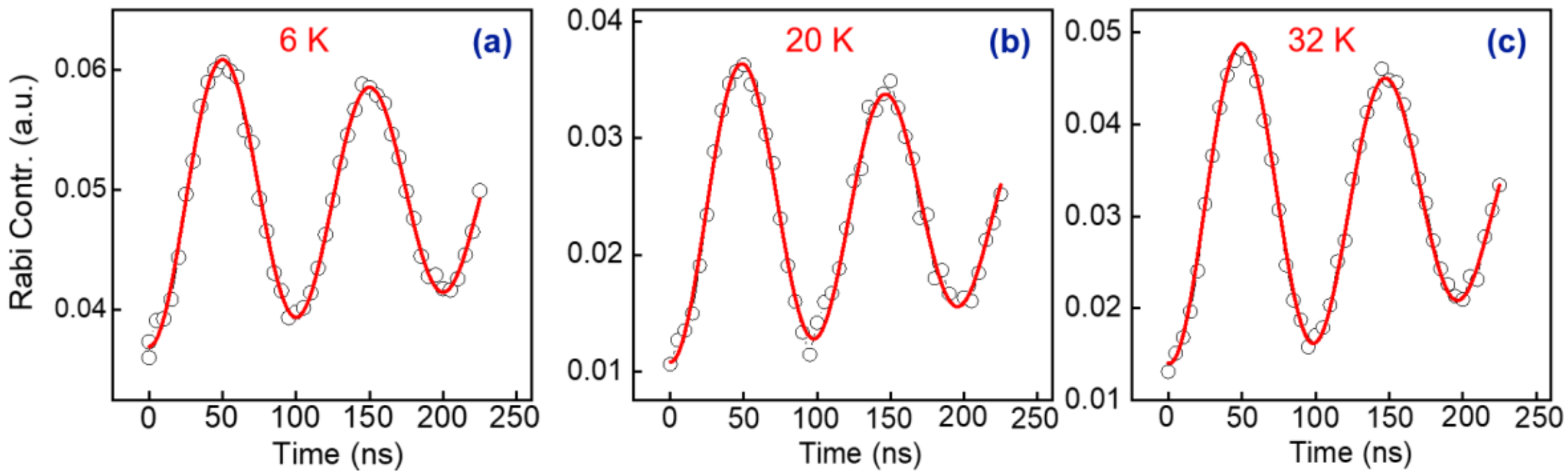

**Figure S4.3:** Rabi curves off the $CrCl_3$ flake vs temperature (6 – 37 K) at 27.77 mT.

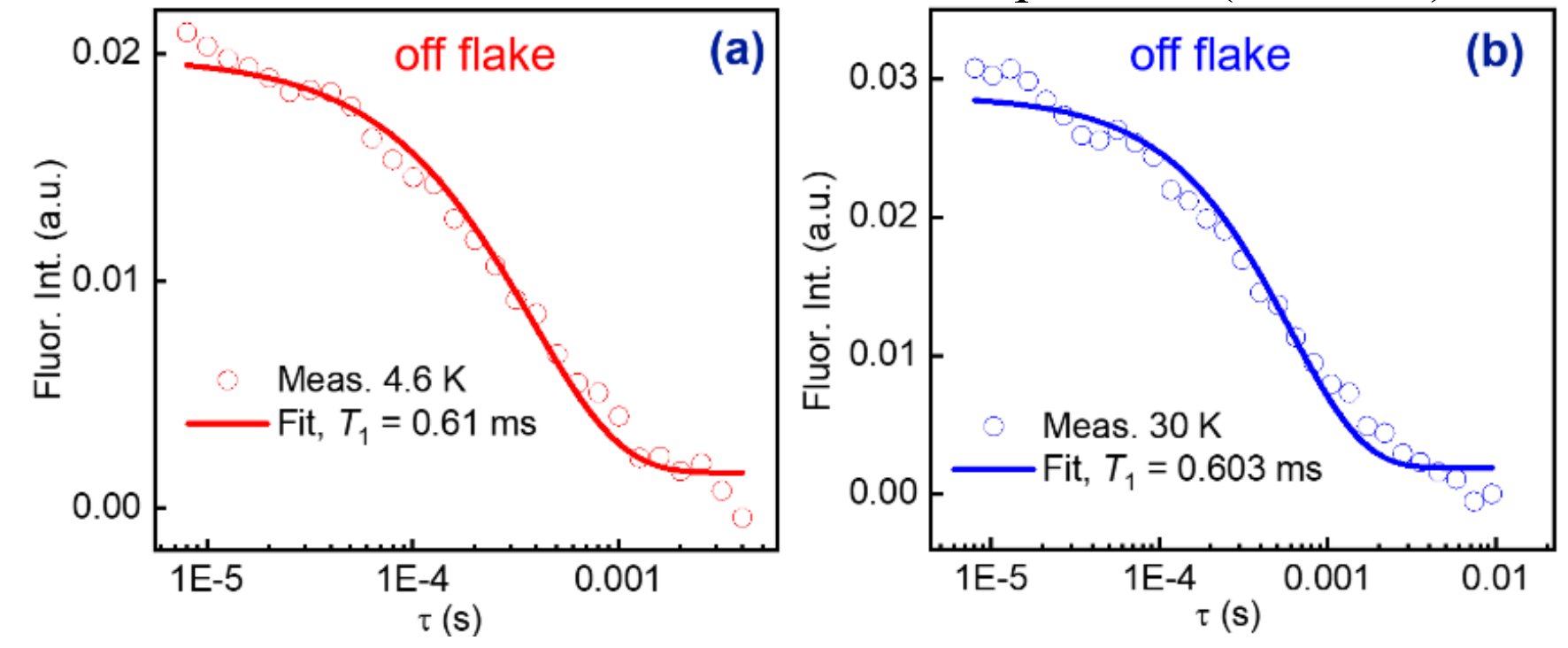

**Figure S4.4:** $T_1$ curves vs temperature (4.6 K and 30 K) off the $CrCl_3$ flake at 27.77 mT.

### *NV Rabi and $T_1$ measurements at 27.77 mT on the 42 nm $CrCl_3$ flake*

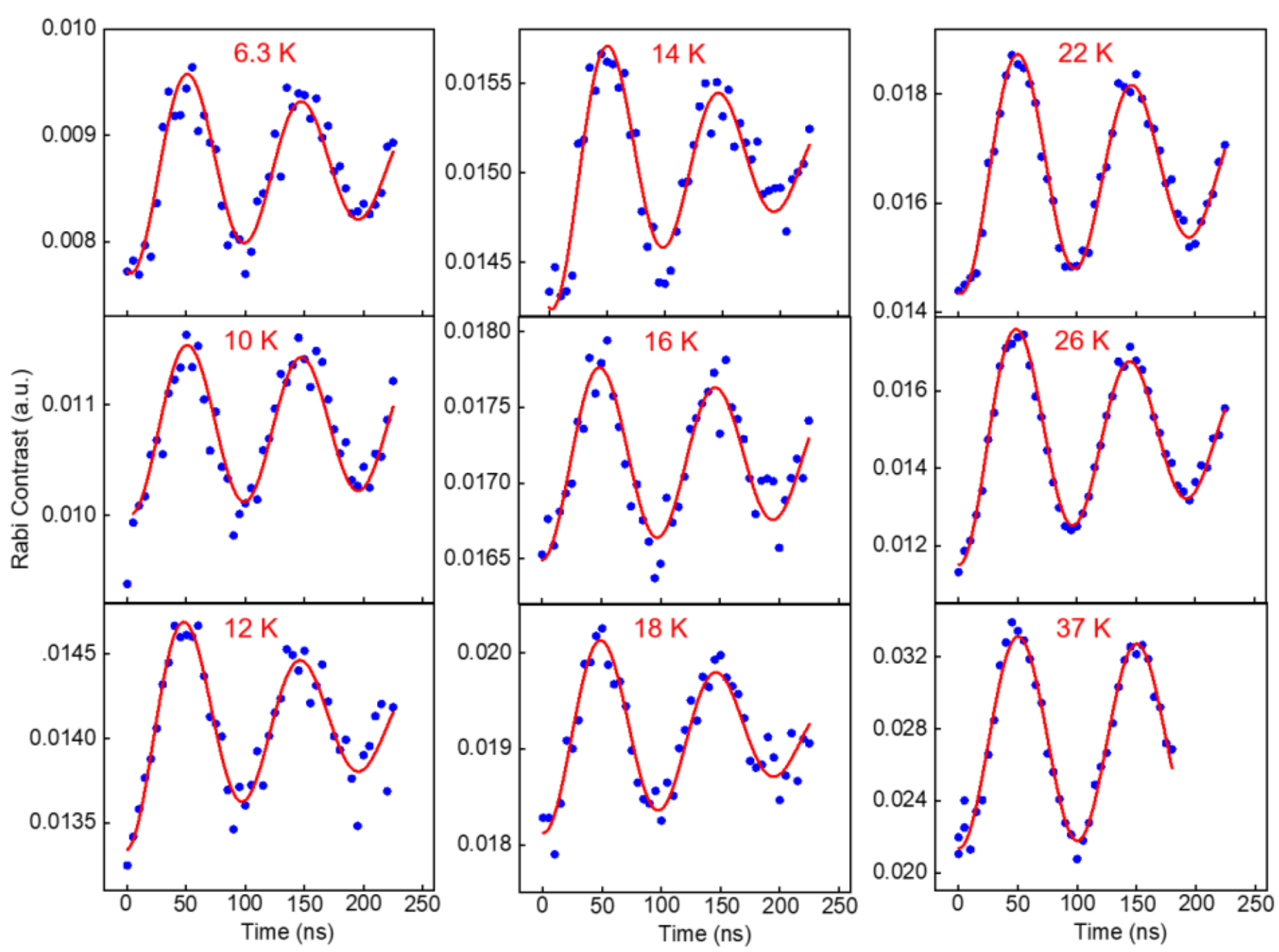

**Figure S4.5:** Rabi curves on the 42 nm $CrCl_3$ flake vs temperature (6.3 – 37 K) at 27.77 mT.

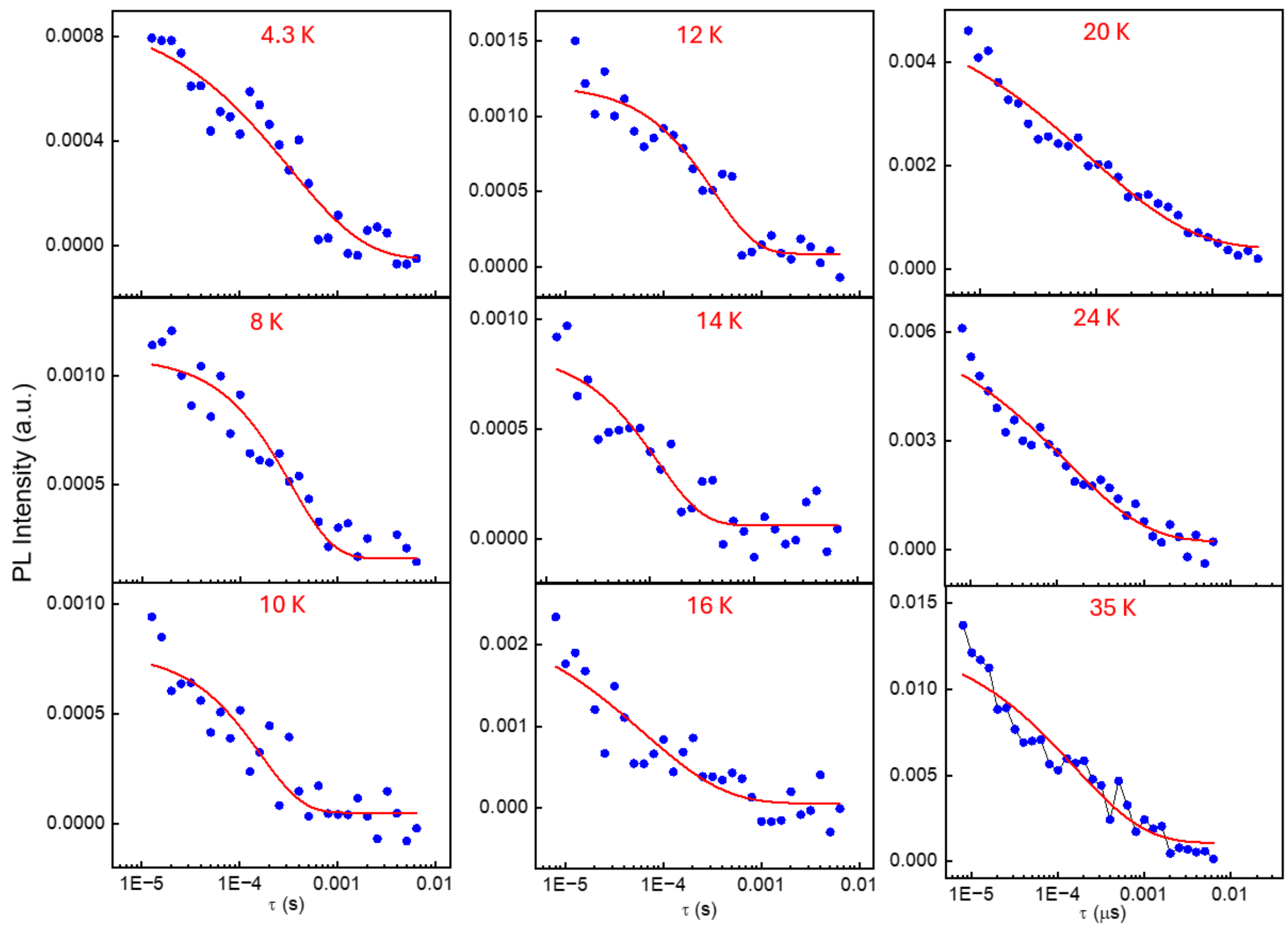

**Figure S4.6:** $T_1$ curves vs temperature (4.3 – 35 K) on the 42 nm $CrCl_3$ flake at 27.77 mT.

## S5. Additional FMR measurements on $CrCl_3$ crystals

Broadband magnetic resonance measurements were performed at 14 K and 23.5 K on the $CrCl_3$ crystals. Figures S5.1a and S5.1c show representative spectra measured at 5 GHz, which are fitted with a single derivative Lorentzian function. The extracted resonance fields were used to construct the frequency–field dispersion, shown in Figure S5.1b, which is fitted using the Kittel equation. The inset displays the linewidth ($\Delta H$) versus frequency (inset of Figure S5.1b), linearly fitted as:[15] $\Delta H = 4\pi\alpha f/\gamma\mu_0 + \Delta H_0$, where α is the damping constant and $\Delta H_0$ is the zero-field (intercept) linewidth of ~16.8 mT. We find α of $\sim 10^{-3}$. Figure S5.1d compares the frequency–field dispersion at 14 K and 23.5 K, where the linear dependence with zero intercept at 23.5 K confirms the paramagnetic phase, and deviation from linear dependence at 14 K confirms the ferromagnetic phase.

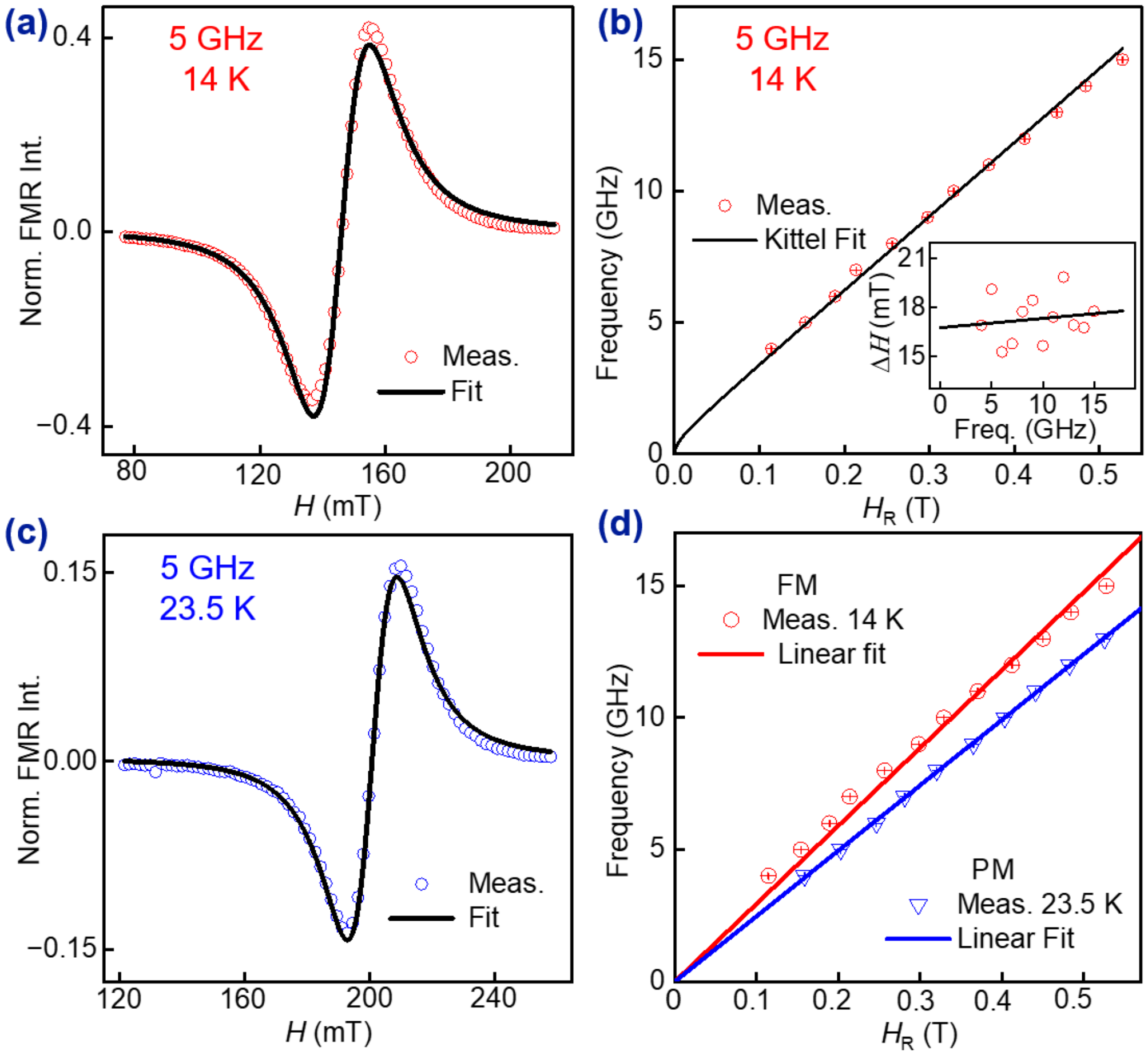

**Figure S5.1:** FMR measurements on $CrCl_3$ crystals at a frequency of 5 GHz at 14 K (a) and 23.5 K (c). (b) FMR frequency versus resonance field at 5 GHz and 14 K. Inset of (b) shows the linewidth ($\Delta H$) as a function of frequency. (d) Measured (scattered lines) FMR frequency versus resonance field at 5 GHz and temperatures of 14 K and 23.5 K.

Additional temperature-dependent magnetic resonance measurements were performed at frequencies of 7 GHz and 10 GHz. Figures S5.2a and S5.2c show the corresponding spectra, which are fitted with the derivative of the Lorentzian function to extract the resonance ($H_R$) field and linewidth ($\Delta H$). The extracted resonance field $H_R$ and linewidth $\Delta H$ are plotted in Figures S5.2b and S5.2d, respectively. Both figures show similar temperature dependence at 5 GHz, indicating that the evolution of the resonance field and linewidth is consistent across the different operating frequency ranges.

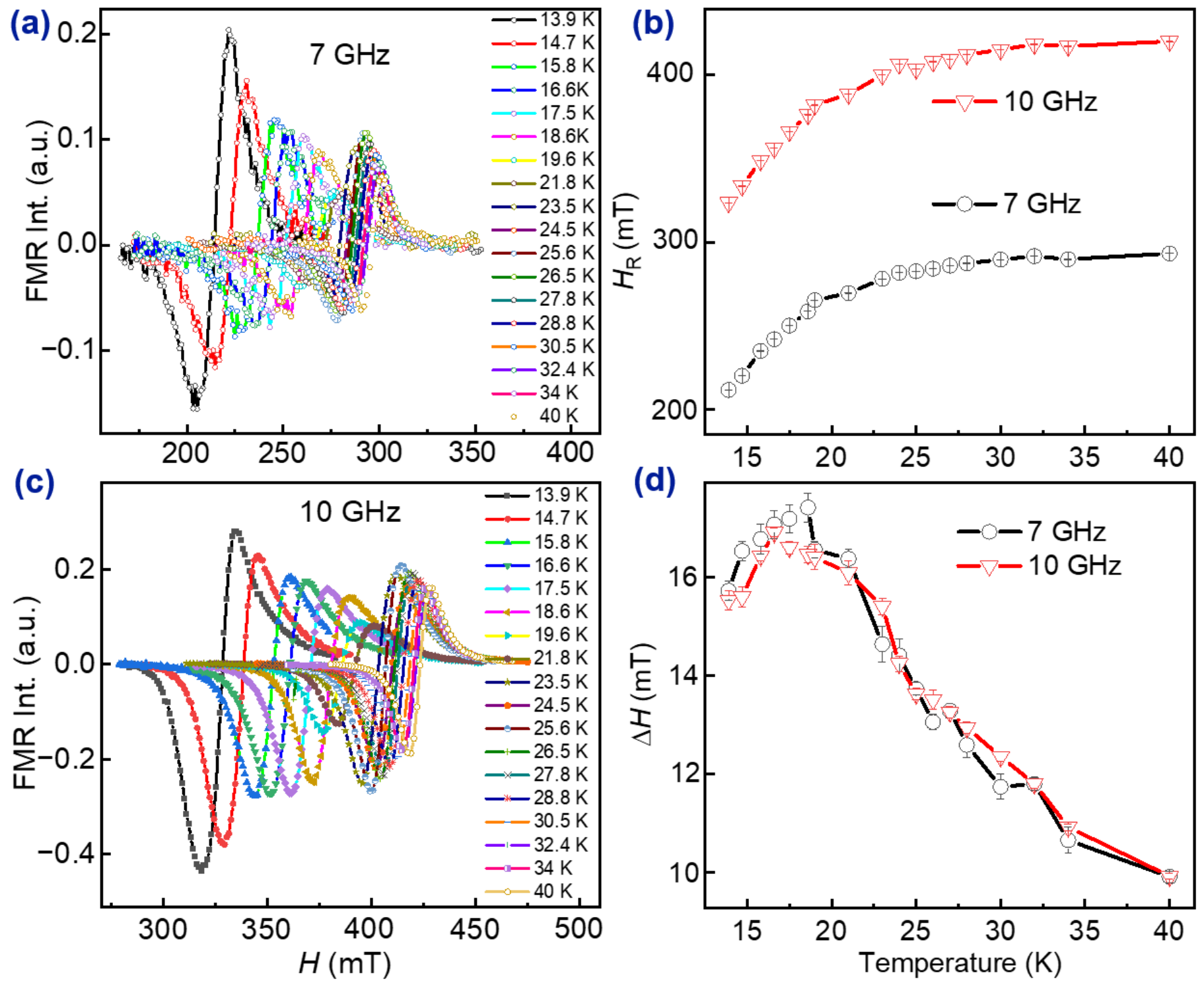

**Figure S5.2:** Temperature-dependent broadband magnetic resonance measurements at 7 GHz and 10 GHz. (a) and (c) Magnetic resonance spectra measured as a function of temperature at 7 GHz and 10 GHz, respectively. (b) Resonance field $H_R$ as a function of temperature extracted from Lorentzian fits of the spectra. (d) Corresponding linewidth $\Delta H$ versus temperature for 7 and 10 GHz frequencies.

In Figure S5.3, we plot the measured FMR (frequency vs resonance magnetic field $H_R$) curves for temperatures in the range of 15.8 – 40 K.

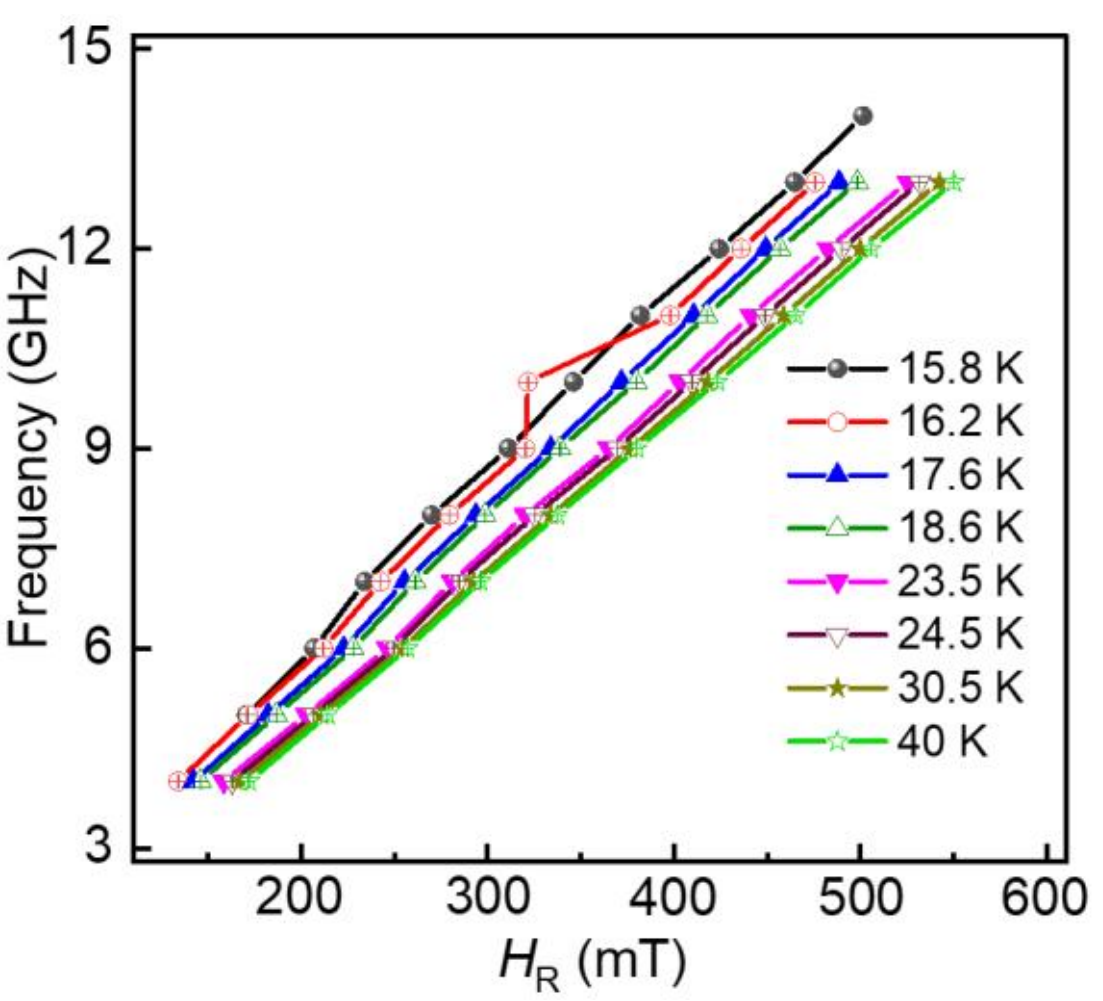

**Figure S5.3:** Frequency–field dispersion obtained from broadband magnetic resonance measurements of $CrCl_3$ at temperatures ranging from 15.8 K to 40 K. The resonance frequency is plotted as a function of the resonance field $H_R$, showing a linear dependence characteristic of paramagnetic resonance above 18 K. It exhibits a ferromagnetic resonance behavior Below 18 K.